\documentclass[a4paper,11pt]{article}
\pdfoutput=1 
\usepackage{physics}

\usepackage{jheppub} 
\usepackage[T1]{fontenc} 
\usepackage{slashed}
\newcommand{\be}{\begin{equation}}
\newcommand{\ee}{\end{equation}}

\usepackage{subcaption}
\usepackage{ucs}
\usepackage{hyperref}
\usepackage[utf8x]{inputenc}
\usepackage{amsmath,bm}
\usepackage{amsfonts}
\usepackage{comment}
\usepackage{amssymb}
\usepackage{array}
\usepackage{epsfig}
\usepackage{slashed}
\usepackage{lipsum}
\usepackage{hyperref}
\hypersetup{
	colorlinks=true,
	linkcolor=blue,
	filecolor=red,
	urlcolor=blue,
	citecolor=red
} 

\title{\boldmath  Double copy structure of parity-violating CFT correlators}

\author[a]{Sachin Jain,}
\author[a]{ Renjan Rajan John,}
\author[a]{Abhishek Mehta,}
\author[b]{Amin A.~Nizami,}
\author[a]{Adithya Suresh}

\affiliation[a]{Indian Institute of Science Education and Research, Homi Bhabha Road, Pashan, Pune 411 008, India}
\affiliation[b]{Department of Physics, Ashoka University, India}

\emailAdd{sachin.jain@iiserpune.ac.in}
\emailAdd{renjan.john@acads.iiserpune.ac.in}
\emailAdd{aan27cam@gmail.com}
\emailAdd{\{abhishek.mehta,s.adithya\}@students.iiserpune.ac.in}

\abstract{We show that general parity-violating 3d conformal field theories show a double copy structure for momentum space 3-point functions of conserved currents, stress tensor and marginal scalar operators. Splitting up the CFT correlator into two parts - called homogeneous and non-homogeneous - we show that double copy relations exist for each part separately. We arrive at similar conclusions regarding double copy structures using tree-level correlators of massless fields in $dS_4$. We also discuss the flat space limit of these correlators.
We further extend the double copy analysis to correlators involving higher-spin conserved currents, which suggests that the spin-$s$ current correlator can be thought of as $s$ copies of the spin one current correlator. }
\begin{document}
	
\maketitle

\raggedbottom
\flushbottom

\section{Introduction}
There has been a remarkable confluence in the study of CFT correlators and scattering amplitudes in recent years.  As is well known, scattering amplitudes can be extracted by taking a suitable limit of appropriate CFT correlators - in position, momentum or Mellin space \cite{Gary:2009ae, Gary:2009mi, Komatsu:2020sag, Penedones:2010ue, Raju:2012zr, Fitzpatrick:2011hu}. This enables a CFT derivation of various flat space amplitude results \cite{Fitzpatrick:2011hu, Fitzpatrick:2011dm}.  Conversely, amplitude methods have recently been used in the study of CFTs \cite{Caron-Huot:2017vep, Gillioz:2020mdd}.


One of the interesting relationships that exists for flat space scattering amplitudes is the double-copy relation between gauge theory and gravity amplitudes, and the associated color-kinematics duality \cite{Kawai:1985xq,Bern:2008qj, Bern:2010ue}. Here, substitution of color factors by kinematic factors in the numerators generates gravity amplitudes from gauge amplitudes, thereby manifesting a quadratic relationship between these two theories. 
This means that amplitudes involving gravitons can be built out from those involving gluons. The double copy relation was first observed in Einstein gravity and pure Yang-Mills theory, and later it was extended to a whole host of theories including higher derivative conformal gravity, higher derivative gauge theories and bi-adjoint scalar theories \cite{Broedel:2012rc, Johansson:2017srf, Johansson:2018ues}. The 3-point structure for the higher-derivative theories is significant because they occur in the momentum space form of CFT correlators of conserved currents, stress tensors and scalars.
Double copy relations also exist for higher point tree amplitudes and loop amplitudes \cite{Bern:2013yya, He:2017spx}.  For a comprehensive review see \cite{ Bern:2019prr}. The analyses in these works were for the parity-even sector. We will show in this paper that similar relationships between amplitudes continue to hold with the inclusion of possible parity-violating terms in the Lagrangian.


In this work we study 3-point CFT correlators in momentum space. Some recent works where momentum space CFTs have been studied include \cite{Coriano:2013jba,Bzowski:2013sza,Bonora:2015nqa,Bonora:2015odi,Bonora:2016ida,sissathesis,Bzowski:2015pba,Bzowski:2015yxv,Bzowski:2017poo,Coriano:2018bbe,Bzowski:2018fql,Gillioz:2018mto,Coriano:2018tgn,Albayrak:2018tam,Farrow:2018yni,Isono:2018rrb, Isono:2019wex,Isono:2019ihz,Maglio:2019grh,Gillioz:2019lgs,Bzowski:2019kwd,Gillioz:2019iye,Bautista:2019qxj,Coriano:2019nkw,Lipstein:2019mpu,Bzowski:2020kfw,Jain:2020rmw,Jain:2020puw,Coriano:2020ccb,Albayrak:2020fyp,Armstrong:2020woi,Serino:2020pyu,Coriano:2020ees,Mata:2012bx,Ghosh:2014kba,Kundu:2014gxa,Arkani-Hamed:2015bza,Maldacena:2011nz, Arkani-Hamed:2018kmz,Baumann:2019oyu,Baumann:2020dch,Skvortsov:2018uru,Jain:2021wyn}.
In particular, the double-copy structures of certain parity-even 3-point functions were inferred in momentum space in \cite{Farrow:2018yni,Lipstein:2019mpu}. In three dimensions, in addition to the parity-even structures, one also needs to consider parity-odd correlators. The most general form of  3-point functions in 3d CFTs is known to be of the form :
\begin{equation}
\langle J_{s_1} J_{s_2} J_{s_3}\rangle = \langle J_{s_1} J_{s_2} J_{s_3}\rangle_{\text{even}}+  \langle J_{s_1} J_{s_2} J_{s_3}\rangle_{\text{odd}}
\end{equation} 
where $J_s$ is a spin $s$ conserved current. In position space one can show that the parity-even part can be obtained by adding contributions arising from the free-boson and the free-fermion theories \cite{Maldacena:2011jn,Maldacena:2012sf,Giombi:2011rz}. However, interacting 3d CFTs such as Chern-Simons-Matter theories can contain a non-trivial parity-odd sector as well \cite{Aharony:2011jz,Giombi:2011kc,Aharony:2012nh,Giombi_2017,Chowdhury:2017vel,Skvortsov_2019}.

In this paper we will demonstrate double copy relations between general parity-violating $CFT_3$ 3-point correlators involving marginal scalars, spin one and spin two conserved currents. We will also show that a double-copy like structure exists for correlators involving higher spin conserved currents. 

To establish our claim, it is convenient to split up CFT correlators into two parts, namely homogeneous and non-homogeneous parts. Their definition will be made clear in the next section. In particular, we show that under double copy relations, the homogeneous part maps to homogeneous part and the non-homogeneous part maps to non-homogeneous part. Let us illustrate this point by considering $\langle TTT \rangle$, the 3-point function of the stress tensor and $\langle JJJ\rangle$ the 3-point function of the conserved spin-1 current.
The correlators can be written as :
\begin{equation}\begin{split}
\langle JJJ\rangle&=  \langle JJJ\rangle_{\text{homogeneous}}+ \langle JJJ\rangle_{\text{non-homogeneous}}\\
\langle TTT\rangle&=  \langle TTT\rangle_{\text{homogeneous}}+ \langle TTT \rangle_{\text{non-homogeneous}}
\end{split}
\end{equation}
The double copy relation is then given by :
\begin{equation}\begin{split}
&\langle TTT\rangle_{\text{homogeneous}}\propto  \left(\langle JJJ\rangle_{\text{homogeneous}}\right)^2\\
&\langle TTT\rangle_{\text{non-homogeneous}}\propto  \left(\langle JJJ\rangle_{\text{non-homogeneous}}\right)^2
\end{split}
\end{equation}
where the proportionality factor is momentum dependent and is different for the two cases. It is given explicitly in Section \ref{jTjT}.

We will show that  double copy relations hold even with the inclusion of the parity violating contributions. We demonstrate this using the results for parity even CFT correlators from \cite{Bzowski:2013sza,Bzowski:2017poo,Bzowski:2018fql,Farrow:2018yni} and parity odd correlators from \cite{Jain:2021wyn,wip} where CFT correlators were obtained by solving conformal Ward identities. There is another interesting way to fix the form of these correlators, initiated in \cite{Maldacena:2011nz}. Here, late-time tree level boundary correlators in Lorentzian $dS_4$ can be computed by first doing an equivalent calculation in flat Minkowski space. Thereafter, using certain conformal properties of the fields in $dS_4$, the corresponding $dS_4$ correlators are obtained by simply dressing the result with a conformal-time integral factor. This Lorentzian $dS_4$  correlator also naturally computes a Euclidean $CFT_3$ correlator. 
We use this method to independently derive the parity-odd structures for the CFT correlators. This method provides a route to obtaining general parity-violating momentum space $CFT_3$ correlators without solving conformal Ward identities \cite{Jain:2021wyn,wip}.

The rest of the paper is organised as follows. In Section \ref{notation}, we introduce the notation used in this paper. In Section \ref{cftcorrelators}, we give the form of all the relevant CFT correlators. In section \ref{DCCFTcorr} we study double copy relations between various pairs of correlators. In Section \ref{dS}, we discuss the flat space limit and write down CFT correlators in terms of tree-level amplitudes without energy conservation for general parity-violating theories of gravitons, gluons and massless scalars in four dimensions. We also discuss here double copy relations for tree level $dS_4$ correlators and flat space scattering amplitudes. In Section \ref{discussion} we conclude and give some directions for future study.
In Appendix \ref{SHvariablesapp} we give our results for the scattering amplitudes and the CFT correlators in terms of the spinor helicity variables. We give details of a few correlators such as $\langle JJT\rangle$ in Appendix \ref{tjjapp}. In Appendix \ref{DCapp} we give the calculational details used in establishing double copy relations. In Appendix \ref{higherspin-correlator} we give explicit momentum space results for a few correlators involving higher spin currents.


\section{Notation and conventions}
\label{notation}
In this paper we denote 4-dimensional Lorentzian momenta and polarisation vectors by $k_i^{\mu}$ and $z_i^{\mu}$ respectively. Here $i$ is a particle index and $\mu = 0,1,2,3$ is the Lorentz index. For massless spin 2 particles the polarisation tensor can be written as an outer product $z_i^{\mu \nu}=z_i ^{\mu}z_i^{\nu}$. We choose the following gauge to work with null momenta :

\begin{align}
\label{kdef}
k_i^{\mu}=(k_i, \vec{k}_i), \,\,\,\, z_i^{\mu}=(0,\vec{z}_i)
\end{align}
where $k_i= |\vec{k}_i|$ is the magnitude of the 3-momentum. 

The 3-dimensional CFT will be Euclidean and current conservation constraints translate to transversality: $k_i\cdot z_i =0$. We will also take $z_i\cdot z_i=0$ which in Euclidean signature implies that the components of $\vec{z}_i$ will be complex.

In our computation we will find it useful to introduce the following notation for various combinations of magnitudes of momenta : 
\begin{align}
E=k_1+k_2+k_3,\quad b_{ij}=k_i k_j,\quad b_{123}=k_1k_2+k_2k_3+k_3k_1,\quad c_{123}=k_1 k_2 k_3
\end{align}
We also introduce the following notation :
\begin{align}
J^2=(k_1+k_2+k_3)(-k_1+k_2+k_3)(k_1-k_2+k_3)(k_1+k_2-k_3)
\end{align}
We will make use of spinor-helicity notation. The momentum vector $p_{\mu}$ for massless scattering in 4-dimensional flat space-time can be written as $p_{\mu}\sigma^{\mu}_{\alpha \dot{\alpha}}= p_{\alpha \dot{\alpha}}= \lambda_{\alpha} \widetilde{\lambda}_{\dot{\alpha}}$ where $\lambda$ denotes a spinor-helicity variable. Since 4d amplitudes are related to 3d CFT correlators it will be useful to have a 3d version of this formalism by utilising the time-like vector $\tau^{\mu}=(1,0,0,0)$, or $\tau^{\alpha \dot{\beta}}=\epsilon^{\alpha \dot{\beta}}$ which can be used to go from dotted to undotted indices (see appendix B of \cite{Farrow:2018yni} and \cite{Lipstein:2019mpu}). We use this to define $\bar{\lambda}^{\alpha}\equiv\tau^{\alpha \dot{\beta}} \widetilde{\lambda}_{\dot{\beta}}$.

For a correlator comprising conserved currents  $J_{s_i}$ the conformal Ward identity in spinor-helicity notation takes the following form :
\begin{align}
\widetilde K^\kappa\left\langle\frac{J_{s_1}}{k_1^{s_{1}-1}}\frac{J_{s_2}}{k_2^{s_{2}-1}}\frac{J_{s_3}}{k_3^{s_{3}-1}}\right\rangle=\text{transverse Ward identity}
\end{align}
where $J_{s_i}$ are conserved currents with spin $s_i$ and dimension $\Delta=s_i+1$ and the R.H.S of the above equation is proportional to the transverse Ward identities associated with the correlator. For instance, for the case of $\langle JJJ\rangle$ where $J$ is the spin-1 conserved current, the conformal Ward identity takes the form :
\begin{align}\label{Kkappajjj}
\widetilde{K}^{\kappa} \langle J^- J^- J^- \rangle &= 2\left(z_1^{-\kappa}\frac{k_{1\mu}}{k_1^2}\langle J^{\mu} J^- J^- \rangle+z_2^{-\kappa}\frac{k_{2\mu}}{k_2^2}\langle J^- J^{\mu} J^- \rangle+z_3^{-\kappa}\frac{k_{3\mu}}{k_3^2}\langle J^- J^- J^{\mu} \rangle\right)
\end{align}
where we see that the R.H.S of the equation is given by the transverse Ward identities.

The correlator $\langle J_{s_1} J_{s_2}J_{s_3}\rangle$ is given by the sum of two terms that satisfy the homogeneous and non homogeneous equations respectively :
\begin{align}
\langle J_{s_1} J_{s_2}J_{s_3}\rangle=\langle J_{s_1} J_{s_2}J_{s_3}\rangle_{\text{homogeneous}}+\langle J_{s_1} J_{s_2}J_{s_3}\rangle_{\text{non homogeneous}}
\end{align}
where $\langle J_{s_1} J_{s_2}J_{s_3}\rangle_{\text{homogeneous}}$ satisfies :
\begin{align}
\widetilde K^\kappa\left\langle\frac{J_{s_1}}{k_1^{s_{1}-1}}\frac{J_{s_2}}{k_2^{s_{2}-1}}\frac{J_{s_3}}{k_3^{s_{3}-1}}\right\rangle_{\text{homogeneous}}=0
\end{align}
and $\langle J_{s_1} J_{s_2}J_{s_3}\rangle_{\text{non homogeneous}}$ satisfies :
\begin{align}
\widetilde K^\kappa\left\langle\frac{J_{s_1}}{k_1^{s_{1}-1}}\frac{J_{s_2}}{k_2^{s_{2}-1}}\frac{J_{s_3}}{k_3^{s_{3}-1}}\right\rangle_{\text{non homogeneous}}=\text{transverse Ward identity}
\end{align}
The transverse Ward identity is determined in terms of two-point functions. Hence in the momentum space expression for the correlator $\langle J_{s_1} J_{s_2}J_{s_3}\rangle$, we identify the part proportional to the 2-point function coefficient to be the solution to the non-homogeneous Ward identity and the part obtained by setting the coefficient of the 2-point function to zero to be the solution to the homogeneous Ward identity. We will use subscripts \textbf{h} and \textbf{nh} to denote the solutions to the homogeneous and non-homogeneous equations respectively.

For a correlator with at least one scalar operator $O_\Delta$ with conformal dimension $\Delta$ the conformal Ward identity has a trivial R.H.S and is given by : 
\begin{align}
\widetilde K^\kappa\left\langle\frac{O_{\Delta}}{k_1^{\Delta-2}}\frac{J_{s_2}}{k_2^{s_{2}-1}}\frac{J_{s_3}}{k_3^{s_{3}-1}}\right\rangle=0
\end{align}
which holds true both when $s_2=s_3$ and $s_2\ne s_3$.

We will denote flat space amplitudes by $\mathcal{A}$. The corresponding correlators in $dS_4$ \footnote{up to overall conformal time integral factors as discussed in section 5}, with all insertions at the equal-time spatial-slice $\eta=0$, will be denoted by $\mathcal{M}$.

\section{CFT correlators} \label{cftcorrelators}
In this section we present the momentum space expressions for 3-point $CFT_3$ correlators comprising spin-1 conserved current $J$, stress tensor $T$, higher spin conserved currents $J_s$ with spin $s>2$  and marginal scalar operators $O_3$. The parity-even sector of 3-point CFT correlators has been studied by solving the associated conformal Ward identities in a series of works \cite{Bzowski:2013sza,Bzowski:2017poo,Bzowski:2018fql,Bzowski:2015pba}. In \cite{Jain:2021wyn,wip} we studied the parity-odd sector of 3-point correlators by solving conformal Ward identities and using the technique of spin-raising and weight-shifting operators in momentum space \cite{Karateev:2017jgd,Baumann:2019oyu,Baumann:2020dch}. 

We present our results after contracting the momentum space expressions with null transverse polarization vectors.

\subsection{$\langle J_sJ_sO_3\rangle$ for general spin $s$ current}
In this subsection we write down correlators of the form $\langle J_sJ_sO_3\rangle$ for a general spin $s$. For $s=1,2$ we can write down their explicit form in momentum space easily. A similar momentum space expression for a correlator involving general spin $s$ current is very cumbersome. However, it takes a very simple form when written in terms of spinor-helicity variables. We also note that, as discussed in the previous section, a correlator of the form $\langle J_sJ_sO_3\rangle$ only has a homogeneous part. The homogeneous correlator is given by the contribution from the parity-even and the parity-odd sectors :
\begin{equation}
\ev{J_s J_s O_3}_{\text{\bf{h}}} =\ev{J_s J_s O_3}_{\text{even,\bf{h}}}  + \ev{J_s J_s O_3}_{\text{odd,\bf{h}}}  
\end{equation}
\subsubsection{$\langle JJO_3\rangle$}
Let us first consider the 3-point correlator comprising two spin-1 conserved currents and a marginal scalar operator. The momentum space expression for the parity-even part of the correlation function is \cite{Bzowski:2013sza,Bzowski:2018fql,Farrow:2018yni} :
\be\begin{split}
\label{jjo3evencft}
\ev{JJO_3}_{\text{even,\bf{h}}}&=\frac{(E+k_3) }{E^2} \left[2(\vec{z}_1\cdot \vec{k}_2)( \vec{z}_2\cdot \vec{k}_1) +E (E-2k_3)\vec{z}_1\cdot \vec{z}_2 \right]\\
\ev{JJO_3}_{\text{even,\bf{nh}}}&=0.
\end{split}\ee
The momentum space expression for the parity-odd part of the correlator is \cite{Jain:2021wyn,wip} :
\begin{align}
\label{jjo3oddcft}
\langle JJO_3\rangle_{\text{odd,\bf{h}}}&= \frac{(E+k_3) }{E^2}\left[k_2\,\epsilon^{ k_1 z_1 z_2}- k_1\,\epsilon^{k_2 z_1  z_2}\right]\nonumber\\
\ev{JJO_3}_{\text{odd,\bf{nh}}}&=0.
\end{align}
%

\subsubsection{$\langle TTO_3\rangle$}
Let us now consider the 3-point correlator comprising two stress-tensor insertions and a marginal scalar. %
The parity-even part of the correlator is given by  \cite{Bzowski:2013sza,Bzowski:2018fql,Farrow:2018yni} :
\begin{align}
\label{tto3evencft}
\ev{TTO_3}_{\text{even,\bf{h}}}
&= k_1 k_2\frac{E+3k_3}{E^4} \left[ 2 (\vec{z}_1\cdot \vec{k}_2)( \vec{z}_2\cdot \vec{k}_1) +E (E-2k_3)\vec{z}_1\cdot \vec{z}_2 \right]^2\cr
\ev{TTO_3}_{\text{even,\bf{nh}}}&=0
\end{align}
whereas the parity-odd part of this correlator is \cite{wip} :
\begin{align}
\label{tto3oddcft}
\langle TTO_3\rangle_{\text{odd,\bf{h}}}
&=\frac{E+3k_3}{E^4} (k_2 \epsilon^{ k_1 z_1 z_2}- k_1 \epsilon^{k_2 z_1  z_2})\cr
&\hspace{2cm}\times (\vec{z}_1 \cdot \vec{z}_2)(\vec{k}_1\cdot\vec{k}_2-k_1 k_2)\-(\vec{z}_1\cdot \vec{k}_2)( \vec{z}_2\cdot \vec{k}_1)\cr
\ev{TTO_3}_{\text{odd,\bf{nh}}}&=0
\end{align}

\subsubsection{$\langle J_sJ_sO_3\rangle$} Correlators of the form $\langle J_sJ_sO_3\rangle$ have unwieldy expressions in momentum space. The easiest way to determine them is to use weight-shifting operators \cite{Baumann:2020dch}. 
For this, we first derive the momentum space expressions for correlators with spin-3 and spin-4 currents using weight-shifting operators \cite{Baumann:2020dch} :
\begin{equation}\label{s34}
\begin{split}
\langle J_3 J_3 O_3\rangle_{\text{even,\bf{h}}}&= P^{(3)}_1 P^{(3)}_2 H_{12}^3 \langle O_2 O_2 O_3 \rangle\\
\langle J_4 J_4 O_3\rangle_{\text{even,\bf{h}}}&= P^{(4)}_1 P^{(4)}_2 H_{12}^4 \langle O_2 O_2 O_3 \rangle
\end{split}
\end{equation}
where $P^{(s)}_i$ are spin-$s$ projectors \cite{Baumann:2020dch} and $H_{12}$ is a bi-local weight shifting operator which raises the spin at points 1 and 2 and lowers the dimensions at points 1 and 2. In momentum space the operator takes the form \cite{Baumann:2020dch} :
\begin{align}
H_{12}=2(\vec z_1\cdot\vec K_{12})(\vec z_2\cdot\vec K_{12})-(\vec z_1\cdot\vec z_{2})K_{12}^2
\end{align}
where $K_{12}^i\equiv\frac{\partial}{\partial k_1^i}-\frac{\partial}{\partial k_2^i}$.
The explicit form of the correlator in \eqref{s34} is complicated and not reproduced here. A similar expression can be written down for parity-odd contribution. 

These correlators when expressed in spinor-helicity variables take a very simple form. For this, let us consider the 3-point correlator of two higher spin conserved currents $J_s$ with spin $s$ and a marginal scalar. %
The parity-even part of the correlator is given by  \cite{wip} :
\begin{align}
\label{jsjso3evencft}
\begin{split}
\langle J^{-}_{s}J^{-}_{s}O_3\rangle_{\text{even,\bf{h}}}&= \frac{E+(2s-1)k_3}{E^{2s}}\langle 12 \rangle^{2s}\\[5 pt]
\langle J^{+}_{s}J^{+}_{s}O_3\rangle_{\text{even,\bf{h}}}&= \frac{E+(2s-1)k_3}{E^{2s}}\langle \overline{12} \rangle^{2s}
\end{split}
\end{align}
whereas the parity-odd part of the correlator is \cite{wip} :
\begin{align}
\label{jsjso3oddcft}
\begin{split}
\langle J^{-}_{s}J^{-}_{s}O_3\rangle_{\text{odd,\bf{h}}}= i\frac{E+(2s-1)k_3}{E^{2s}}\langle 12 \rangle^{2s}\\[5 pt]
\langle J^{+}_{s}J^{+}_{s}O_3\rangle_{\text{odd,\bf{h}}}= -i\frac{E+(2s-1)k_3}{E^{2s}}\langle \overline{12} \rangle^{2s}
\end{split}
\end{align}
As stated earlier, the non-homogeneous piece vanishes for both parity-even and parity-odd correlators :
\begin{equation}
\langle J_sJ_sO_3\rangle_{\bf{nh}}=0.
\end{equation}
One can check that for the cases when $s=1$ and $s=2$, the momentum space results in \eqref{jjo3evencft}, \eqref{jjo3oddcft}, \eqref{tto3evencft}, and \eqref{tto3oddcft} when re-expressed in spinor-helicity variables match the above results.


\subsection{$\langle J_sJ_sJ_s\rangle$ }
In this subsection we write down correlators of the form $\langle J_sJ_sJ_s\rangle$ for a general $s$. Unlike $\langle J_sJ_sO\rangle$, these correlators have both homogeneous and non-homogeneous pieces :
\begin{equation}
\ev{J_s J_s J_s}_{\text{\bf{h}}} =\ev{J_s J_s J_s}_{\text{even,\bf{h}}}  + \ev{J_s J_s J_s}_{\text{odd,\bf{h}}}.  
\end{equation}
We give the explicit forms for $\langle JJJ\rangle$ and $\langle TTT\rangle$ in momentum space. In both cases, there are exactly two homogeneous pieces, one parity-even and another parity-odd. We also note that there is only one parity-even non-homogeneous contribution. The non-homogeneous part of the parity-odd correlator is always a contact term. For a general spin $s$, the momentum space expression is very complicated. However, in spinor-helicity variables it becomes simple and we express the homogeneous part in these variables. The analogous expression for non-homogeneous part is not yet known. 

\subsubsection{$\langle JJJ\rangle$ }
Let us now consider the 3-point correlator comprising three spin-1 conserved current insertions. The parity-even part of the correlator is given by \cite{Bzowski:2013sza,Bzowski:2017poo,Farrow:2018yni} :
\begin{align}\label{jjjevencft}
\ev{JJJ}_{\text{even,\bf{h}}}&= \frac{c_{JJJ}^{\text{even}}}{E^3} \Big[2\,(\vec{z}_1\cdot \vec{k}_2) \, (\vec{z}_2\cdot \vec{k}_3) \, (\vec{z}_3\cdot \vec{k}_1)+E \{k_3\, (\vec{z}_1\cdot \vec{z}_2) \, (\vec{z}_3\cdot \vec{k}_1)+ \text{cyclic}\}\Big]\nonumber\\[5pt]
\ev{JJJ}_{\text{even,\bf{nh}}}&= - \frac{2c_{JJ}^{\text{even}}}{E} [(\vec{z}_1\cdot \vec{z}_2)(\vec{z}_3\cdot \vec{k}_1)+ \text{cyclic}]
\end{align}
Note that $c_{JJ}^{\text{even}}$ appears in two point function  $\ev{J_{\mu}(k)J_{\nu}(-k)}_{\text{even}}=c_{JJ}^{\text{even}} \pi_{\mu\nu}(k) k$ where $ \pi_{\mu\nu}(k)$ is the transverse projector.
The parity-odd part of the correlator is given by \cite{Jain:2021wyn,wip}
\begin{align}\label{jjjoddcft}
\langle JJJ\rangle_{\text{odd,\bf{h}}}&=\frac{c_{JJJ}^{\text{odd}}}{E^3}\left[\left\{(\vec{k}_1 \cdot \vec{z}_3)\left(\epsilon^{k_3 z_1 z_2}k_1-\epsilon^{k_1 z_1 z_2}k_3\right)+(\vec{k}_3 \cdot \vec{z}_2)\left(\epsilon^{k_1 z_1 z_3}k_2-\epsilon^{k_2 z_1 z_3}k_1\right)\right.\right.\nonumber\\[5 pt]
&\hspace{1.5cm}\left.\left.-(\vec{z}_2 \cdot \vec{z}_3)\epsilon^{k_1 k_2 z_1}E+\frac{k_1}{2} \epsilon^{z_1 z_2 z_3}E(E-2k_1)\right\}+\text{cyclic perm}\right]\nonumber\\[5 pt]
\langle JJJ\rangle_{\text{odd,\bf{nh}}}&=c_{JJ}^{\text{odd}} \epsilon^{z_1z_2z_3}
\end{align}
where $c_{JJ}^{\text{odd}}$ arises in parity-odd contribution to the two point function $\ev{J_{\mu}(k)J_{\nu}(-k)}_{\text{odd}}=c_{JJ}^{\text{odd}} \epsilon_{\mu\nu k} .$
Let us also note that the non-homogeneous contribution to the parity-odd part of $\langle JJJ\rangle$
(term proportional to $c_{JJ}^{\text{odd}} $) in \eqref{jjjoddcft} is a contact term.

\subsubsection{$\langle TTT\rangle$}
\label{tttsectioncft}
Let us now consider the 3-point correlator comprising three stress-tensor insertions. The parity-even contribution to the correlator is given by \cite{Bzowski:2013sza,Bzowski:2017poo,Farrow:2018yni} :
\begin{align}\label{tttevencft}
\ev{TTT}_{\text{even,\bf{h}}}
&=\frac{c_{TTT}^{\text{even}}c_{123}^2}{J^2E^5} \Big[(\vec{z}_1\cdot \vec{k}_2 \, \vec{z}_2\cdot \vec{k}_3 \, \vec{z}_3\cdot \vec{k}_1)^2 +\frac{E}{2}\,\vec{z}_1\cdot \vec{k}_2 \, \vec{z}_2\cdot \vec{k}_3 \, \vec{z}_3\cdot \vec{k}_1\,\,  (k_3\, \vec{z}_1\cdot \vec{z}_2 \, \vec{z}_3\cdot \vec{k}_1+ \text{cyclic}) \Big] \nonumber\\[5pt]
\ev{TTT}_{\text{even,\bf{nh}}}&= 2 c_{TT}^{\text{even}} \Big[  \Big(\frac{c_{123}}{E^2} + \frac{b_{123}}{E}- E  \Big)(\vec{z}_1\cdot \vec{z}_2\, \vec{z}_3\cdot \vec{k}_1+ \text{cyclic})^2  +(k_1^3+k_2^3+k_3^3)\mathcal{A}_{ct}\Big]
\end{align}
where $c_{TT}^{\text{even}}$ appears in the even part of the 2-point function $\langle TT\rangle$ and  
\begin{equation}
\begin{split}
\mathcal{A}_{ct}&= (\vec{z}_1\cdot \vec{z}_2)(\vec{z}_2\cdot \vec{z}_3)(\vec{z}_3\cdot \vec{z}_1).
\end{split}\end{equation}
We note that the term proportional to $\mathcal{A}_{ct}$ in \eqref{tttevencft} is a contact term and will be ignored below in establishing the double copy relation.

We see that apart from the physical pole at $E=0$, $\ev{TTT}_{\text{even,\bf{h}}}$ displays an unphysical pole at $J=0$. However, this is only an artefact of the basis we have chosen to work with and by working in a suitable basis we can get rid of the unphysical pole. For instance, it can be shown that after a clever use of 3d degeneracies, one can express the correlator as follows\footnote{See section \ref{CFTds4} for a derivation using gravity.} 
\begin{align}\label{tttevennewbasis}
\ev{TTT}_{\text{even,\bf{h}}}
&=\frac{c_{TTT}^{\text{even}}\,c_{123}}{E^6}F_2[1,2,3]F_2[2,3,1]F_2[3,1,2]
\end{align}
where
\begin{align}
F_2[i,j,l]=\left[(\vec{z}_i\cdot \vec{z}_j)E(E-2k_l)+2 (\vec{z}_i\cdot \vec{k}_j)(\vec{z}_j\cdot \vec{k}_i)\right].
\end{align}
For details of this computation see Appendix \ref{mf3mw3appendix}.

%

The parity-odd contribution to the correlator is given by \cite{wip}
\begin{align}
\label{tttoddcft}
\begin{split}
\langle TTT \rangle_{\text{odd}}&= A_1 \epsilon^{k_3 k_1 z_1}\epsilon^{k_1 k_2 z_2}\epsilon^{k_2 k_3 z_3}(\vec k_2 \cdot\vec  z_1)(\vec k_3 \cdot \vec z_2)(\vec k_1 \cdot\vec  z_3)+ A_2 \epsilon^{k_2 k_3 z_3}(\vec k_1 \cdot \vec z_3)(\vec k_3 \cdot\vec  z_2)^2(\vec k_2 \cdot \vec z_1)^2\\[5 pt]
&+ A_2 (k_2 \leftrightarrow k_3)\epsilon^{k_1 k_2 z_2}(\vec  k_1 \cdot\vec  z_3)^2(\vec k_3 \cdot \vec z_2)(\vec k_2 \cdot\vec  z_1)^2+ A_2 (k_1 \leftrightarrow k_3)\epsilon^{k_3 k_1 z_1}(\vec k_1 \cdot\vec  z_3)^2(\vec k_3 \cdot\vec  z_2)^2(\vec k_2 \cdot\vec  z_1)
\end{split}
\end{align}
where the homogeneous piece in the form factor is given by :
\begin{align}\label{formfac}
A_{1,\bf{h}} &=c_{TTT}^{\text{odd}}\frac{c_{123}^2}{2J^4 E^4},~~~
A_{2,\bf{h}} = -c_{TTT}^{\text{odd}}\frac{b_{12} c^2_{123}}{2J^4 E^4}
\end{align}
Just as in the parity-even case we see that the form factors have an unphysical extra pole at $J=0$. This can again be gotten rid of by working in a suitable basis
where it takes the form\footnote{See section \ref{CFTds4}.} 
\begin{align}
\langle TTT \rangle_{\text{odd}}
&=c_{TTT}^{\text{odd}} ~ \frac{c_{123}}{E^6}\left[2(\vec z_1\cdot\vec  k_2\,\vec z_2\cdot\vec k_3\,\vec z_3\cdot \vec k_1)+E\left((\vec z_1\cdot \vec z_2\,\vec z_3\cdot\vec  k_1)k_3+\text{cyclic}\right)\right]\notag\\[5pt]
&\hspace{.5cm}\left[-(\vec{k}_3 \cdot \vec{z}_2)\left(\epsilon^{k_3 z_1 z_3}k_1+\epsilon^{k_1 z_1 z_3}(E-k_3)\right)-(\vec{k}_1 \cdot \vec{z}_3)\left(\epsilon^{k_2 z_1 z_2}k_1+\epsilon^{k_1 z_1 z_2}(E-k_2)\right)\right.\notag\\[5 pt]
&\hspace{.5cm}\left.+E(\vec{z}_2 \cdot \vec{z}_3)\epsilon^{k_1 k_2 z_1}-\frac{1}{2}k_1E(E-2k_1)\epsilon^{z_1 z_2 z_3}+\text{cyclic perm.}\right]
\end{align}

The non-homogeneous part of the parity-odd correlator is determined by the form factors in \eqref{tttoddcft}
\begin{align}
A_{1,\bf{nh}}  &=-c_{TT}^{\text{odd}} \frac{12(k_1^2+k_2^2+k_3^2)}{J^4}\\[5 pt]
A_{2,\bf{nh}}  &= c_{TT}^{\text{odd}}\frac{\left(k_3^4+7k_3^2(k_1^2+k_2^2)+4(k_1^4+4k_1^2 k_2^2+k_2^4)\right)}{J^4}
\end{align}
where $c_{TT}^{\text{odd}}$ appears in the parity-odd part of the two point function $\langle TT\rangle.$ It can be shown that, the parity-odd non-homogeneous contribution to the correlator is a contact term, see \cite{wip} for details. Since the non-homogeneous part is a contact term, this will be ignored while checking double copy relations.

\subsubsection{$\langle J_sJ_sJ_s\rangle$}
Let us now consider the 3-point correlator comprising three higher spin conserved currents $J_s$ with spin $s$. %
The parity-even contribution to the correlator is given by \cite{wip} :
\begin{align}
\label{jsjsjsevencft}
\begin{split}
\ev{J^{-}_{s}J^{-}_{s}J^{-}_{s}}_{\text{even,\bf{h}}}= \frac{(k_1 k_2 k_3)^{s-1}}{E^{3s}}\langle 12 \rangle^s \langle 23 \rangle^s \langle 31 \rangle^s\\[5 pt]
\ev{J^{+}_{s}J^{+}_{s}J^{+}_{s}}_{\text{even,\bf{h}}}= \frac{(k_1 k_2 k_3)^{s-1}}{E^{3s}}\langle \overline{12} \rangle^s \langle \overline{23} \rangle^s \langle \overline{31} \rangle^s
\end{split}
\end{align}
The parity-odd contribution to the correlator is given\footnote{Let us note that the parity-odd correlator, when written in spinor-helicity variables, gets an extra factor of $i$  as compared to parity-even correlator. This arises due to the fact that when converting momentum space parity-odd answer to spinor-helicity variables, the Levi-Civita tensor $\epsilon^{ijk}$ is expressed in terms of Pauli matrices and their algebra gives rise to this extra factor. More precisely
$$Tr(\sigma_i \sigma_j \sigma_k)=2 i \epsilon_{ijk}.$$} by \cite{wip} 
\begin{align}
\label{jsjsjsoddcft}
\begin{split}
\ev{J^{-}_{s}J^{-}_{s}J^{-}_{s}}_{\text{odd,\bf{h}}}= i \frac{(k_1 k_2 k_3)^{s-1}}{E^{3s}}\langle 12 \rangle^s \langle 23 \rangle^s \langle 31 \rangle^s\\[5 pt]
\ev{J^{+}_{s}J^{+}_{s}J^{+}_{s}}_{\text{odd,\bf{h}}}= -i \frac{(k_1 k_2 k_3)^{s-1}}{E^{3s}}\langle \overline{12} \rangle^s \langle \overline{23} \rangle^s \langle \overline{31} \rangle^s
\end{split}
\end{align}

\section{Double copy structure of CFT correlators}
\label{DCCFTcorr}
In this section we discuss the double copy structure of CFT 3-point correlation functions in momentum space. 
We will see that the double copy relations are such that homogeneous terms are mapped to homogeneous terms and non-homogeneous terms are mapped to non-homogeneous terms.  We will establish our claims in momentum space for $\langle TTO_3\rangle$ and $\langle JJO_3\rangle$ and for  $\langle TTT\rangle$  and $\langle JJJ\rangle$. We use the spinor-helicity variables to show this for correlators such as  $\langle J_s J_s O_3\rangle$ and $\langle J_s J_s J_s\rangle$.


\subsection{$\langle TTO_3\rangle$ and $\langle JJO_3\rangle$}
The following double copy structure of $\langle TTO_3\rangle_{\text{even}}$ was established in \cite{Farrow:2018yni} :
\begin{align}\label{dcpy3}
\langle TTO_3\rangle_{\text{even,\bf{h}}}=\frac{(E+3k_3)k_1k_2}{(E+k_3)^2}\langle JJO_3\rangle_{\text{even,\bf{h}}}\langle JJO_3\rangle_{\text{even,\bf{h}}}
\end{align}
 From the explicit expressions for the correlators in \eqref{jjo3evencft}, \eqref{jjo3oddcft}  and \eqref{tto3oddcft} we notice that the double copy relations extends to the parity-odd sector :
\begin{align}\label{dcpy1}
\langle TTO_3\rangle_{\text{odd,\bf{h}}}=\frac{(E+3k_3)k_1k_2}{(E+k_3)^2}\langle JJO_3\rangle_{\text{odd,\bf{h}}}\langle JJO_3\rangle_{\text{even,\bf{h}}}
\end{align}
Remarkably we also notice from \eqref{jjo3oddcft} and \eqref{tto3evencft} that  $\langle TTO_3\rangle_{\text{even}}$ is also given by the square of $\langle JJO_3\rangle_{\text{odd}}$ 
\begin{align}\label{dcpy2}
\langle TTO_3\rangle_{\text{even,\bf{h}}}=\frac{(E+3k_3)k_1k_2}{(E+k_3)^2}\langle JJO_3\rangle_{\text{odd,\bf{h}}}\langle JJO_3\rangle_{\text{odd,\bf{h}}}
\end{align}
The above double copy relations for $\langle TTO_3\rangle_{\text{even}}$ and $\langle TTO_3\rangle_{\text{odd}}$ immediately imply the following double copy structure for the complete correlator :
\begin{align}\label{dcpyTJo}
\langle TTO_3\rangle_{\text{even,\bf{h}}}+  \langle TTO_3\rangle_{\text{odd,\bf{h}}} &=\frac{(E+3k_3)k_1k_2}{(E+k_3)^2}\left( \langle JJO_3\rangle_{\text{even,\bf{h}}}+  \langle JJO_3\rangle_{\text{odd,\bf{h}}}\right)^2\nonumber\\
\implies\langle TTO_3\rangle_{\text{\bf{h}}}&=\frac{(E+3k_3)k_1k_2}{(E+k_3)^2}\langle JJO_3\rangle^2_{\text{\bf{h}}}
\end{align}
In writing the above double copy relation it is crucial that we have the following relation between $\langle JJO_3\rangle_{\text{even}}$ and $\langle JJO_3\rangle_{\text{odd}}$ :
\begin{align}
\langle JJO_3\rangle_{\text{even,\bf{h}}}^2=c\langle JJO_3\rangle_{\text{odd,\bf{h}}}^2
\end{align}
where $c$ is some constant.
As noted in Section \ref{notation}, for correlators such as $\langle TTO_3\rangle$ and $\langle JJO_3\rangle$ the conformal Ward identity (in spinor helicity variables) does not have a non-homogeneous term. Hence the double copy structure that we obtained above is purely for correlators that satisfy the homogeneous conformal Ward identity.

\subsection{Double copy for higher-spin correlators}\label{HS}
 We will now extend our analysis of the double copy structure of $\langle TTO_3\rangle$ to higher spin correlators of the form $\langle J_sJ_sO_3\rangle$. 
%
%
 Using \eqref{s34} one can show that the higher spin correlators take the following form :
\begin{equation}
\begin{split}\label{dc4321}
\langle J_4 J_4 O_3\rangle_{\text{even,\bf{h}}}&= k_1 k_2 \frac{(E+7k_3)}{(E+3k_3)^2}\langle J_2 J_2 O_3\rangle_{\text{even}}^2\\
\langle J_3 J_3 O_3\rangle_{\text{even,\bf{h}}}&=k_1k_2\frac{(E+5k_3)}{(E+k_3)(E+3k_3)}\langle J_2 J_2 O_3\rangle_{\text{even}}\langle J_1 J_1 O_3\rangle_{\text{even}}.
\end{split}
\end{equation}
Calculating the parity-odd contribution to these three point functions is difficult due to the high amount of degeneracy \cite{Jain:2021wyn,wip}. However, in spinor helicity variables the computation becomes easier. In these variables one has the following remarkable relation between the parity-even and parity-odd contributions \cite{wip} :
\begin{equation}\label{oddevenr}
\begin{split}
\langle {J_s}^{-}{J_s}^{-} O_3\rangle_{\text{even},\bf{h}} &= i \langle {J_s}^{-}{J_s}^{-} O_3\rangle_{\text{odd},\bf{h}}, ~~\langle {J_s}^{+}{J_s}^{+} O_3\rangle_{\text{even},\bf{h}} =- i \langle {J_s}^{+}{J_s}^{+} O_3\rangle_{\text{odd},\bf{h}} 
\end{split}
\end{equation}
for any $s$ and all the other spinor helicity components are zero. Using this one can generalise \eqref{dc4321} to include the parity-odd sector. 
The double copy relation \eqref{dc4321} then becomes
\begin{equation}\begin{split}\label{dc4321a}
\langle J_4 J_4 O_3\rangle_{\text{even},\bf{h}}+ \langle J_4 J_4 O_3\rangle_{\text{odd},\bf{h}}&= \frac{ k_1 k_2(E+7k_3)}{(E+3k_3)^2}\left(\langle J_2 J_2 O_3\rangle_{\text{even},\bf{h}}+ \langle J_2 J_2 O_3\rangle_{\text{odd},\bf{h}}\right)^2\nonumber\\[5pt]
\implies\langle J_4 J_4 O_3\rangle_{\bf{h}}&= \frac{ k_1 k_2(E+7k_3)}{(E+3k_3)^2}\langle J_2 J_2 O_3\rangle_{\bf{h}}^2\nonumber\\[5pt]
\langle J_3 J_3 O_3\rangle_{\text{even},\bf{h}}+ \langle J_3 J_3 O_3\rangle_{\text{odd},\bf{h}}&=\frac{k_1k_2(E+5k_3)}{(E+k_3)(E+3k_3)}\left(\langle J_2 J_2 O_3\rangle_{\text{even},\bf{h}}+ \langle J_2 J_2 O_3\rangle_{\text{odd},\bf{h}} \right)\nonumber\\&\hspace{4cm}\times\left(\langle J_1 J_1 O_3\rangle_{\text{even},\bf{h}}+ \langle J_1 J_1 O_3\rangle_{\text{odd},\bf{h}} \right)\nonumber\\[5pt]
\implies\langle J_3 J_3 O_3\rangle_{\bf{h}}&=\frac{k_1k_2(E+5k_3)}{(E+k_3)(E+3k_3)}\langle J_2 J_2 O_3\rangle_{\bf{h}}\langle J_1 J_1 O_3\rangle_{\bf{h}}
\end{split}
\end{equation}
To write down the above double copy relations it is crucial that we have the following relation between the parity-odd and parity-even parts of the correlators :
\begin{align}
\langle J_2 J_2 O_3\rangle_{\text{even},\bf{h}}^2&=\langle J_2 J_2 O_3\rangle_{\text{odd},\bf{h}}^2\nonumber\\[5pt]
\langle J_2 J_2 O_3\rangle_{\text{even},\bf{h}} \langle J_1 J_1 O_3\rangle_{\text{even},\bf{h}}&=\langle J_2 J_2 O_3\rangle_{\text{odd},\bf{h}} \langle J_1 J_1 O_3\rangle_{\text{odd},\bf{h}}
\end{align}
As we noted in the case of the double copy structure of $\langle TTO_3\rangle$ in terms of $\langle JJO_3\rangle$ the conformal Ward identity for correlators of the form $\langle J_sJ_sO_3\rangle$  does not have a non-homogeneous term. Hence the double copy relations that we arrived at here are purely for the homogeneous terms.

\subsection*{Double copy relation for general spin}
One can easily extend the above analysis to correlators of the form $\langle J_sJ_sO_3\rangle$ in the spinor helicity variables, see \cite{wip}. The correlation functions are given by :
\begin{equation}\begin{split}
\langle J_{s}^-J_{s}^- O \rangle_{\text{even},\bf{h}} &=k_1^{s-1}k_2^{s-1}\frac{(E+(2s-1)k_3)}{E^{2s}}\langle 12 \rangle^{2s}\\
\langle J_{s}^+J_{s}^+ O \rangle_{\text{even},\bf{h}} &=k_1^{s-1}k_2^{s-1}\frac{(E+(2s-1)k_3)}{E^{2s}}\langle \bar{1}\bar{2} \rangle^{2s}\\
\langle J_{s}^+J_{s}^- O \rangle_{\text{even},\bf{h}} &=\langle J_{s}^-J_{s}^+ O \rangle_{\text{even},\bf{h}} =0.
\end{split}\end{equation}
The spinor helicity components of the odd part of the correlator can be computed using \eqref{oddevenr}.
One can then derive the following double copy relation for a general correlator of the kind $\langle J_sJ_sO_3\rangle$ that satisfies the homogeneous conformal Ward identity :
\begin{equation}\begin{split}\label{dcssp}
&\langle J_s J_s O\rangle_{\text{even},\bf{h}}+ \langle J_s J_s O\rangle_{\text{odd},\bf{h}}\nonumber\\
&=\frac{k_1 k_2(E+(2s-1)k_3)}{(E+(2s'-1)k_3)(E+(2s''-1)k_3)}\left(\langle J_{s'} J_{s'} O\rangle_{\text{even},\bf{h}}+ \langle J_{s'} J_{s'}O\rangle_{\text{odd},\bf{h}} \right)\nonumber\\ 
&\hspace{6.5cm}\times\left(\langle J_{s''} J_{s''} O\rangle_{\text{even},\bf{h}}+ \langle J_{s''} J_{s''} O\rangle_{\text{odd},\bf{h}} \right)\nonumber\\
&\implies\langle J_s J_s O\rangle_{\bf{h}}=\frac{k_1 k_2(E+(2s-1)k_3)}{(E+(2s'-1)k_3)(E+(2s''-1)k_3)}\langle J_{s'} J_{s'} O\rangle_{\bf{h}}\langle J_{s''} J_{s''} O\rangle_{\bf{h}}
\end{split}\end{equation}
where $s'+s''=s$.

We will now come to more complicated correlators such as $\langle JJJ\rangle$ and $\langle TTT\rangle $ whose conformal Ward identities have a non-homogeneous term and show more interesting double copy relations.
\subsection{Double copy relation for $\langle JJJ\rangle$ and $\langle TTT\rangle $}\label{jTjT}
The double copy relation between $\langle JJJ\rangle$ and $\langle TTT\rangle$ is more subtle than those for correlators with a scalar operator insertion. Unlike $\langle TTO_3\rangle$ or $\langle JJO_3\rangle$ these correlators have a non-homogeneous term as well and we will see that the double copy structures map homogeneous terms to homogeneous terms and non-homogeneous terms get mapped to non-homogeneous terms.

\subsection*{Homogeneous terms}
The following double copy structure was noticed in \cite{Farrow:2018yni} for the homogeneous term in the even part of $\langle TTT\rangle$ :
\begin{align}
\langle TTT\rangle_{\text{even},\bf{h}}=k_1k_2k_3\langle JJJ\rangle_{\text{even},\bf{h}}\langle JJJ\rangle_{\text{even},\bf{h}}
\end{align}
From the explicit expressions for the correlators in \eqref{jjjevencft}, \eqref{jjjoddcft}, and \eqref{tttoddcft}, we notice :
\begin{align}
\langle TTT\rangle_{\text{odd},\bf{h}}=k_1k_2k_3\langle JJJ\rangle_{\text{odd},\bf{h}}\langle JJJ\rangle_{\text{even},\bf{h}}
\end{align}
We also have the remarkable relation that the parity-even part of the homogeneous term is given by the square of the odd part of the homogeneous term in $\langle JJJ\rangle$ : 
\begin{align}
\langle TTT\rangle_{\text{even},\bf{h}}=k_1k_2k_3\langle JJJ\rangle_{\text{odd},\bf{h}}\langle JJJ\rangle_{\text{odd},\bf{h}}
\end{align}
Combining these relations we obtain the following double copy relation for the complete homogeneous term of the $\langle TTT\rangle$ correlator :
\begin{align}
\langle TTT\rangle_{\text{even},\bf{h}}+\langle TTT\rangle_{\text{odd},\bf{h}}&=k_1k_2k_3(\langle JJJ\rangle_{\text{even},\bf{h}}+\langle JJJ\rangle_{\text{odd},\bf{h}})^2\nonumber\\
\implies \langle TTT\rangle_{\bf{h}}&=k_1k_2k_3\,\langle JJJ\rangle_{\bf{h}}^2
\end{align}
\subsection*{Non-homogeneous terms}
From \eqref{jjjevencft} and \eqref{tttevencft} we know that $\langle JJJ \rangle_{\text{even}}$ and $\langle TTT \rangle_{\text{even}}$ also have non-trivial non-homogeneous parts between which there exists the following double copy relation :
\begin{align}
\langle TTT \rangle_{\text{even,\bf{nh}}} = (E^3 - E(k_1 k_2+k_2 k_3+k_1 k_3)-k_1 k_2 k_3)\langle JJJ \rangle_{\text{even,\bf{nh}}}^2
\end{align}
This relation is independent of the double copy of the homogeneous part as the pre-factor is different. 
The non-homogeneous parts of $\langle TTT \rangle_{\text{odd}}$ and $\langle JJJ \rangle_{\text{odd}}$ are trivial as they are contact terms.

\subsection{Double copy structure for higher spin correlators}
We will now discuss the double copy structures in higher spin correlators. This is most easily done using the spinor-helicity variables. 
The parity-even part of the homogeneous part of $\langle J_{s}J_sJ_s \rangle$ is given by \cite{wip} :
\begin{align}
\langle J_{s}J_sJ_s \rangle_{\text{even},\bf{h}} =\frac{(k_1 k_2 k_3)^{s-1}}{E^{3s}} \langle 12 \rangle^s \langle 23 \rangle^s \langle 31 \rangle^s
\end{align}
As noted in \cite{wip}, the odd part of the above correlator is given by the same expression up to an overall factor of $i$ :
\begin{align}
\langle J_{s}J_sJ_s \rangle_{\text{odd},\bf{h}}  =i\,\frac{(k_1 k_2 k_3)^{s-1}}{E^{3s}} \langle 12 \rangle^s \langle 23 \rangle^s \langle 31 \rangle^s
\end{align}
From this, we have the following double copy expression for the homogeneous part of the higher spin correlator $\langle J_{s} J_{s} J_{s} \rangle$ :
\begin{align}
\label{jsjsjsdccft}
\langle J_{s} J_{s} J_{s} \rangle_{\bf{h}} &= k_1 k_2 k_3(\langle J^{s'}J^{s'}J^{s'} \rangle \langle J^{s''}J^{s''}J^{s''} \rangle)
\end{align}
such that $s'+s'' = s$. 

\subsection{Spin $s$ current correlator as $s$ copies of the spin one current correlator}\label{hs1}
In this section we note that we can write correlators of the form $\langle J_sJ_sO_3\rangle$ and $\langle J_sJ_sJ_s\rangle$ as $s$ copies of correlators of the spin-one current. Using the double copy relations in \eqref{dcssp} recursively we notice that :
\begin{align}
\langle J_s J_s O\rangle_{\bf{h}}=(k_1k_2)^{s-1}\,\frac{E+(2s-1)k_3}{(E+k_3)^s}\left(\langle J J O\rangle_{\bf{h}}\right)^s
\end{align}
Similarly using the double copy relations in \eqref{jsjsjsdccft} recursively we notice that :
\begin{align}
\langle J_s J_s J_s\rangle_{\bf{h}}=(k_1k_2k_3)^{s-1}\left(\langle JJJ\rangle_{\bf{h}}\right)^s
\end{align}

\section{ CFT correlators from $dS_4$ Feynman diagrams and the flat space limit} \label{dS}
In this section, we relate CFT correlators discussed in the previous section to the tree-level amplitude calculated using Feynman diagrams in $dS_4$. We also relate CFT correlators to flat space scattering amplitudes.
\subsection{Amplitudes}
Here we will study 3-point flat-space scattering amplitudes in general parity-violating theories of gravitons, gluons and massless scalars in 4d. These are calculated straightforwardly from a Lagrangian whose  cubic vertices will contribute to the 3-point amplitudes. We  will use the notation $\mathcal{M}$ to denote the flat space amplitude without energy conservation. This is a useful quantity because $E=k_1+k_2+k_3 \ne 0$  in $dS_4$ and the $dS_4$ vertex is obtained from $\mathcal{M}$ by multiplying with an overall conformal time integral factor. This also matches the CFT correlators that we computed in Section \ref{cftcorrelators}.
We also take flat space limit of $\mathcal{M}$ and resultant scattering amplitude will be denoted by $\mathcal{A}.$ More precisely
\begin{equation}
\mathcal{A} = \lim_{E \rightarrow 0} \mathcal{M}.
\end{equation}
We also express $\mathcal{M}$  in both 4d and 3d notations. See section \ref{notation}
and the discussion below \eqref{kdef} for how  to express 4d amplitude in terms of three dimensional notations. The 3d notation is particularly useful while comparing with the CFT correlators.

We will first describe the gauge theory action, the gravity action and the gravity-gluon interactions that we consider.

\subsection*{Gauge theory action}

The gauge theory action that we consider is :
\begin{align}
S_A = S_{EM}+S^A_{\text{even}}+S^A_\text{odd}
\end{align}
where $S_{EM}$ is the electromagnetic action and $S^A_{\text{even}}$ and $S^A_\text{odd}$ are gauge invariant parity-preserving and parity-violating actions respectively, given by :
\begin{align}
&S^{EM}_{\text{even}} = -\frac{1}{4}\int d^4x\sqrt{g} F^2\label{ema}\\[5pt]
&S^A_{\text{even}} = \int d^4x\sqrt{g}(\alpha^A_1 F^3+\alpha^A_2 \phi F^2)\label{saeven}\\[5pt]
&S^A_\text{odd} = \int (\beta^A_1F_{\mu\nu}F_{\rho\sigma}+\beta^A_2F_{\mu\nu}F_{\rho}^{~\tau}F_{\sigma\tau}+\beta^A_3\phi F_{\mu\nu}
F_{\rho\sigma})dx^{\mu}\wedge dx^{\nu} \wedge dx^{\rho} \wedge dx^{\sigma}\label{saodd}
\end{align}
where
\begin{align}
F^3 = F_{\alpha}^{~\beta}F_{\beta}^{~\gamma}F_{\gamma}^{~\alpha}
\end{align}
Depending on the background, $g_{\mu \nu}$ will be the four dimensional Minkowski or de Sitter metric.

\subsection*{Gravity action}
The gravity action that we consider is :
\begin{align}
S_g = S_{EH}+S^g_{\text{even}}+S^g_\text{odd}
\end{align}
where $S_{EH}$ is the Einstein-Hilbert action and  $S^g_{\text{even}}$ and $S^g_\text{odd}$ are parity-preserving and parity-violating actions respectively, given by :
\begin{align}
&S_{EH}=\frac{1}{16 \pi G} \int d^4x\sqrt{g} (R+\Lambda)\label{EH}\\[5pt]
&S^g_{\text{even}} = \int d^4x\sqrt{g}(\alpha^g_1 W^2+\alpha^g_2 W^3+\alpha^g_3 \phi W^2)\label{sgeven}\\[5pt]
&S^g_\text{odd} = \int (\beta^g_1W_{\mu\nu\rho\sigma}W_{\alpha\beta}^{~~~\rho\sigma}+\beta^g_2W_{\rho\sigma\alpha\beta}W^{\sigma\gamma}_{~~\mu\tau}W_{\gamma~~~\nu}^{~\rho\tau}+\beta^g_3\phi W_{\mu\nu\rho\sigma}W_{\alpha\beta}^{~~~\rho\sigma})dx^{\mu}\wedge dx^{\nu} \wedge dx^{\alpha} \wedge dx^{\beta}\label{sgodd}
\end{align}
where
\be
W^2 = W_{\mu\nu\rho\sigma}W^{\mu\nu\rho\sigma}, \quad W^3 = W_{\mu\nu\rho\sigma}W^{\rho\sigma\alpha\beta}W_{\alpha\beta\mu\nu}, \quad g= \det(g_{\mu \nu}) 
\ee
and $W_{\mu\nu\rho\sigma}$ is the Weyl tensor.


\subsubsection{Gauge amplitudes}
\subsubsection*{Gluon-gluon-scalar amplitudes}

Let us first compute the contribution to the gluon-gluon-scalar amplitudes due to the interactions corresponding to the coupling $\alpha^A_2$ and $\beta^A_3$ in \eqref{saeven} and \eqref{saodd}.
From the Lagrangian, we first compute the amplitude without energy conservation ($\mathcal{M}$) and express it in both 4d and 3d notation (see \eqref{kdef}) we obtain
\begin{align}\label{fsqandfftilda}
\mathcal{M}_{\phi F^2} &=2(k_1 \cdot z_2)(k_2 \cdot z_1)-2 (k_1 \cdot k_2) (z_1 \cdot z_2)\nonumber\\[5pt]
&=2(\vec{k}_1 \cdot \vec{z}_2)(\vec{k}_2 \cdot \vec{z}_1)+E(E-2k_3)(\vec{z}_1 \cdot \vec{z}_2)\nonumber\\[5pt]
\mathcal{M}_{\phi F \widetilde{F}} &= \epsilon^{z_1 k_1 z_2 k_2}=-\epsilon^{z_1 z_2 k_1}k_2+\epsilon^{z_1 z_2 k_2} k_1
\end{align}
%
%
To get actual flat space scattering amplitudes we have to impose energy conservation, i.e. $E\rightarrow 0$. This gives :
\begin{align}\label{Aphis}
\mathcal{A}_{\phi F^2} &=\frac{1}{2}\lim_{E \rightarrow 0} \mathcal{M}_{\phi F^2} 
=(\vec{k}_1 \cdot \vec{z}_2)(\vec{k}_2 \cdot \vec{z}_1),\quad\nonumber\\
\mathcal{A}_{\phi F \widetilde{F}} &=\lim_{E \rightarrow 0} \mathcal{M}_{\phi F \widetilde{F}} 
=-\epsilon^{z_1 z_2 k_1}k_2+\epsilon^{z_1 z_2 k_2} k_1.
\end{align}

\subsubsection*{Gluon-gluon-gluon amplitudes}

Let us now compute the contribution to the gluon-gluon-gluon amplitudes due to interactions corresponding to couplings $\alpha^A_1$ and $\beta^A_2$ in \eqref{saeven} and \eqref{saodd}. The parity-even amplitude is given by :
%
\begin{align}
\label{mF3F2Ft1}
\mathcal{M}_{F^3}&= z_{[1\mu}k_{1\nu]}z_{[2\nu}k_{2\rho]}z_{[3\rho}k_{3\mu]}\nonumber\\[5 pt]
&=2 (\vec{k}_2 \cdot \vec{z}_1)(\vec{k}_3 \cdot \vec{z}_2)(\vec{k}_1 \cdot \vec{z}_3)+E(k_1(\vec{k}_2 \cdot \vec{z}_1)(\vec{z}_2 \cdot \vec{z}_3)+\text{cyclic perm.})
\end{align}
The parity-odd amplitude is given by :
\begin{align}
\label{mF3F2Ft}
\mathcal{M}_{F^2 \widetilde{F}} &= z_{[2\alpha}k_{2\tau]}z_{[3\tau}k_{3\beta]}\epsilon^{\alpha\beta z_1 k_1}+\text{cyclic perm.}\nonumber\\[5 pt]
&= \left[-(\vec{k}_1 \cdot \vec{z}_3)\left(\epsilon^{k_3 z_1 z_2}k_1-\epsilon^{k_1 z_1 z_2}k_3\right)+(\vec{k}_3 \cdot \vec{z}_2)\left(\epsilon^{k_1 z_1 z_3}k_2-\epsilon^{k_2 z_1 z_3}k_1\right)\right.\nonumber\\[5 pt]
&\hspace{1cm}\left.-(\vec{z}_2 \cdot \vec{z}_3)\epsilon^{k_1 k_2 z_1}E+\frac{k_1}{2} \epsilon^{z_1 z_2 z_3}E(E-2k_1)\right]+\text{cyclic perm.}
\end{align}
The flat space scattering amplitudes is obtained by taking the $E\rightarrow 0$ limit. This gives :
\begin{align}
\label{gluonamps}
\mathcal{A}_{F^3}&=\frac 12\lim_{E\rightarrow 0}\mathcal{M}_{F^3} 
=(\vec{k}_2 \cdot \vec{z}_1)(\vec{k}_3 \cdot \vec{z}_2)(\vec{k}_1 \cdot \vec{z}_3)\nonumber\\[5 pt]
\mathcal{A}_{F^2 \widetilde{F}} 
&=\left[-(\vec{k}_1 \cdot \vec{z}_3)\left(\epsilon^{k_3 z_1 z_2}k_1-\epsilon^{k_1 z_1 z_2}k_3\right)+(\vec{k}_3 \cdot \vec{z}_2)\left(\epsilon^{k_1 z_1 z_3}k_2-\epsilon^{k_2 z_1 z_3}k_1\right)\right]+\text{cyclic perm.}
\end{align}
We also have contributions to the 3-gluon amplitude from $F^2$ and $F\widetilde{F}$. To obtain these we first calculate :
\begin{align}
\label{mF2andmFFt}
\mathcal{M}_{YM} &= (k_2 \cdot z_1)(z_2 \cdot z_3) +\text{cyclic perm.}\nonumber=(\vec{k}_2 \cdot \vec{z}_1)(\vec{z}_2 \cdot \vec{z}_3) +\text{cyclic perm.}\nonumber\\[5pt]
\mathcal{M}_{F\widetilde{F}} &= \epsilon^{k_1 z_1 z_2 z_3}+\text{cyclic perm.}= E\,\epsilon^{z_1 z_2 z_3}
\end{align}
%
The actual flat space scattering amplitudes are given by :
\begin{align}
\label{gluonamps1}
\mathcal{A}_{YM} &= (\vec k_2 \cdot \vec z_1)(\vec z_2 \cdot \vec z_3) +\text{cyclic perm.}\nonumber\\[5 pt]
\mathcal{A}_{F\widetilde{F}} &= 0
\end{align}
\subsubsection{Gravity amplitudes}
Working with a traceless, transverse perturbation around the flat metric, we obtain the following for the Weyl tensor to first order in perturbation :
\begin{align}
W_{\mu\nu\rho\sigma} = k_{[\mu}z_{\nu]}k_{[\rho}z_{\sigma]}
\end{align}
where $k^{\mu}$ and $z^{\mu}$ are defined in \eqref{kdef}. 

\subsubsection*{Graviton-graviton-scalar amplitude}
Let us consider the graviton-graviton-scalar amplitude due to the interaction corresponding to the coupling $\beta^g_3$ in \eqref{sgodd}. From the Lagrangian we have :
\begin{align}
\label{mphiWWt}
\mathcal{M}_{\phi W\widetilde{W}}&=\epsilon^{\alpha\beta\gamma\delta}{W_{(1)}}_{\mu\nu\alpha\beta}{W_{(2)}}^{\mu\nu}_{~~\gamma\delta}\nonumber\\[5pt]
& = \epsilon(z_1k_1z_2k_2)\left(2(k_2 \cdot z_1)(k_1 \cdot z_2)+E(E-2k_3)z_1\cdot z_2\right)\nonumber\\[5 pt]
&= \left(2(\vec{z}_1\cdot \vec{k}_2)(\vec{z}_2 \cdot \vec{k}_1)+E(E-2k_3)\vec{z}_1\cdot \vec{z}_2\right)\left(-k_1\epsilon^{z_1 z_2 k_2}+k_2 \epsilon^{z_1 z_2 k_1}\right)
\end{align}
%
%
The flat space amplitude is given by :
\begin{align}
\label{AphiWWt}
\mathcal{A}_{\phi W\widetilde{W}}
= 2(\vec{z}_1\cdot \vec{k}_2)(\vec{z}_2 \cdot \vec{k}_1)\left(-k_1\epsilon^{z_1 z_2 k_2}+k_2 \epsilon^{z_1 z_2 k_1}\right)
\end{align}
Similarly, one can compute the graviton-graviton-scalar amplitude due to the interaction corresponding to the coupling $\alpha^g_3$ in \eqref{sgeven} from $\mathcal{M}_{\phi W^2}$ :
\begin{align}
\label{mphiWsquared}
\mathcal{M}_{\phi W^2}&={W_{(1)}}_{\mu\nu\rho\sigma}{W_{(2)}}^{\mu\nu\rho\sigma}\nonumber\\[5pt]
& = \left(2(z_1\cdot k_2)(z_2\cdot k_1)+E(E-2k_3)z_1 \cdot z_2\right)^2 \nonumber\\[5 pt]
&= \left(2(\vec{z}_1\cdot \vec{k}_2)(\vec{z}_2\cdot \vec{k}_1)+E(E-2k_3)\vec{z}_1\cdot \vec{z}_2\right)^2 
\end{align}
The flat space amplitude is obtained by taking the $E\rightarrow 0$ limit :
\begin{align}
\label{aphiw2}
\mathcal{A}_{\phi W^2}
= \left(2(\vec{z}_1\cdot \vec{k}_2)(\vec{z}_2\cdot \vec{k}_1)\right)^2 
\end{align}
\subsubsection*{Graviton-graviton-graviton amplitude}
Here we calculate graviton-graviton-graviton amplitude from interactions corresponding to couplings $\alpha^g_2$ and $\beta^g_2$ in \eqref{sgeven} and \eqref{sgodd}. 
 The parity-even  amplitude is given by :
\begin{align}
\mathcal{M}_{W^3} &= F_1(1,2,3)  F_1(2,3,1)F_1(3,1,2)\notag\\
&=  F_2(1,2,3)  F_2(2,3,1)F_2(3,1,2)\label{MWWW}
\end{align}
where
\begin{align}
F_1(i,j,l) &= \left[-2(z_i\cdot z_j)k_i\cdot k_j+2 (z_i\cdot k_j)(z_j\cdot k_i)\right]\cr
F_2(i,j,l) &= \left[(\vec{z}_i\cdot \vec{z}_j)E(E-2k_l)+2 (\vec{z}_i\cdot \vec{k}_j)(\vec{z}_j\cdot \vec{k}_i)\right]
\end{align}
The parity-odd contribution to the amplitude is given by :
\begin{align}
\mathcal{M}_{W^2\widetilde{W}} &= \left(z_{[1\mu}k_{1\nu]}z_{[2\nu}k_{2\rho]}z_{[3\rho}k_{3\mu]}][z_{[2\alpha}k_{2\tau]}z_{[3\tau}k_{3\beta]}\epsilon^{\alpha\beta z_1k_1}\right)\notag\\[5 pt]
&=\left[2(z_1\cdot k_2\,z_2\cdot k_3\,z_3\cdot k_1)+E\left((z_1\cdot z_2\,z_3\cdot k_1)k_3+\text{cyclic}\right)\right]\notag\\[5pt]
&\hspace{.5cm}\left[\{(z_3\cdot k_2)\epsilon(z_2 k_3 z_1 k_1)-(k_2\cdot k_3)\epsilon(z_1z_2z_3k_1)
-(z_2\cdot z_3)\epsilon(z_1k_1k_2k_3)\right. \notag\\[5pt]
&\hspace{.5cm}\left. +(z_2\cdot k_3)\epsilon(k_2z_3z_1k_1)\right\}+\text{cyclic perms.}]\nonumber\\[5 pt]
&=\left[2(\vec z_1\cdot\vec  k_2\,\vec z_2\cdot\vec k_3\,\vec z_3\cdot \vec k_1)+E\left((\vec z_1\cdot \vec z_2\,\vec z_3\cdot\vec  k_1)k_3+\text{cyclic}\right)\right]\notag\\[5pt]
&\hspace{.5cm}\left[(\vec{k}_3 \cdot \vec{z}_2)\left(\epsilon^{k_3 z_1 z_3}k_1+\epsilon^{k_1 z_1 z_3}(E-k_3)\right)+(\vec{k}_1 \cdot \vec{z}_3)\left(\epsilon^{k_2 z_1 z_2}k_1+\epsilon^{k_1 z_1 z_2}(E-k_2)\right)\right.\notag\\[5 pt]
&\hspace{.5cm}\left.-E(\vec{z}_2 \cdot \vec{z}_3)\epsilon^{k_1 k_2 z_1}+\frac{1}{2}k_1E(E-2k_1)\epsilon^{z_1 z_2 z_3}+\text{cyclic perm.}\right]\label{A1}
\end{align}
The flat space amplitude is obtained by taking the $E\rightarrow 0$ limit :
\begin{align}\label{A2}
\mathcal{A}_{W^3} 
&=8 (\vec z_1\cdot\vec  k_2)(\vec z_2\cdot\vec  k_1)\times (\vec z_2\cdot\vec  k_3)(\vec z_3\cdot\vec  k_2)\times (\vec z_3\cdot \vec k_1)(\vec z_1\cdot \vec k_3) \notag\\[5 pt]
\mathcal{A}_{W^2\widetilde{W}} 
&=\left[2(\vec z_1\cdot\vec  k_2\,\vec z_2\cdot\vec k_3\,\vec z_3\cdot \vec k_1)\right]\notag\\[5pt]
&\hspace{.5cm}\left[(\vec{k}_3 \cdot \vec{z}_2)\left(\epsilon^{k_3 z_1 z_3}k_1-\epsilon^{k_1 z_1 z_3}k_3\right)-(\vec{k}_1 \cdot \vec{z}_3)\left(\epsilon^{k_1 z_1 z_2}k_2-\epsilon^{k_2 z_1 z_2}k_1\right)+\text{cyclic perm.}\right]
\end{align}
The EH term in \eqref{EH} gives 
\begin{equation}\label{EHM}
\mathcal{M}_{EH} =\left( (k_2 \cdot z_1)(z_2 \cdot z_3) +\text{cyclic perm.}\nonumber\right)^2=\left((\vec{k}_2 \cdot \vec{z}_1)(\vec{z}_2 \cdot \vec{z}_3) +\text{cyclic perm.}\right)^2.
\end{equation}
In the flat space limit this gives
\begin{equation}\label{EHA}
\mathcal{A}_{EH} =\left((\vec{k}_2 \cdot \vec{z}_1)(\vec{z}_2 \cdot \vec{z}_3) +\text{cyclic perm.}\right)^2.
\end{equation}

\subsection{CFT correlators from $dS_4$}\label{CFTds4}
In this section we will compute  tree level $dS_4$ cosmological correlators using the method of \cite{Maldacena:2011nz} and also including the relevant parity-odd 3-point vertices in the Lagrangian. The calculation with parity-even vertices was done in Appendix A of \cite{Farrow:2018yni}. 

The idea, due to Maldacena and Pimentel \cite{Maldacena:2011nz} , that will be central to our analysis in this section is that certain cosmological correlators in de Sitter can be constructed directly from the corresponding flat-space amplitudes by dressing with conformal time integrals arising from using conformally covariant transformation properties of fields in $dS_4$. These Lorentzian $dS_4$ correlators also compute boundary (Euclidean) $CFT_3$ correlators, thereby establishing a relationship between these three quantities. \footnote{It is of course true that flat space amplitudes can be generated by an appropriate limit of CFT correlators, the non-trivial part is that we can do the converse at least for certain correlators.}

We will work in (Lorentzian) $dS_4$ with the metric:
\be
ds^2=\frac{1}{\eta^2}(-d \eta^2+dx_i^2)
\ee

For calculating $dS_4$ correlators perturbatively, we first look at the on-shell wave-functions for (linearised) free fields, which can be massless scalars, gauge or gravitational perturbations. These kind of fields will suffice to calculate the correlators we will be interested in.  As is well known, for the scalar we have the Bunch Davies mode function $\phi \sim (1-i k \eta)\exp (i k \eta)$, the gauge field solution is a plane wave just like in flat space $A_{\mu} \sim z_{\mu }\exp (i k \eta)$ and  the linearised gravitational perturbation is given by $\gamma_{\mu \nu}  \sim z_{\mu }z_{\nu }(1-i k \eta)\exp (i k \eta)$. It was noted in \cite{Maldacena:2011nz}, that this results in the Weyl tensors for the linearised gravity perturbations about $dS$ and flat backgrounds being conformally related:

\be
W^{\mu}_{(dS)\, \nu \rho \sigma} \Big( \exp (i k \eta)(1-i k \eta)z_{\mu }z_{\nu} \Big) = (-ik\eta)W^{\mu}_{(flat)\, \nu \rho \sigma} \big(\exp (i k \eta)z_{\mu }z_{\nu} \big),
\ee
whereas the gauge field strength is the same in both backgrounds. This means that the $dS$ correlators of interest to us are the same as corresponding flat space amplitudes without energy condervation ($\mathcal{M}$) upto conformal time integrals which are easily evaluated. In this section, we will calculate the contribution to 3-point functions from parity-violating interaction terms in the Lagrangian of the form $F \widetilde{F} \phi$, $W \widetilde{W} \phi$, $F^2 \widetilde{F}$ and $W^2 \widetilde{W}$. We will take all parity-odd two-point functions to be vanishing as this is the case for the $dS_4$ actions considered here.

We will find that our tree-level $dS_4$ computations will generate the different parts of the corresponding $CFT_3$ correlator (both parity-even and odd). Therefore, this perturbative approach provides a simple method of fixing the form of CFT correlators without taking recourse to solving conformal Ward identities.


\subsubsection{$\langle JJO_3\rangle$}
The term in the action which contributes to the odd part of $\langle JJO_3\rangle$ is given by :
\begin{align}
\int \phi F_{\mu\nu}F_{\rho\sigma}~dx^{\mu} \wedge dx^{\nu} \wedge dx^{\rho} \wedge dx^{\sigma}
\end{align}
The tree-level $dS_4$ correlator corresponding to this interaction is given by \footnote{Since the $dS_4$ correlators match $CFT_3$ correlators we will use the notation $\langle JJO_3\rangle$ instead of $\langle \gamma\gamma\phi\rangle$. We will continue to use similar CFT notations for all $dS_4$ correlators in this section. } :
\begin{align}
\label{jjo3cft}
\langle JJO_3\rangle_{\text{odd}}&=\text{Im}\left[\int_{-\infty}^0 d\eta (1-ik_3 \eta)e^{i\eta E}\right]\mathcal{M}_{\phi F\widetilde{F}} = \frac{E+k_3}{E^2}\mathcal{M}_{\phi F\widetilde{F}} 
\end{align}
Substituting for $\mathcal{M}_{\phi F\widetilde{F}}$ from \eqref{fsqandfftilda} we see that this matches the expression for the homogeneous part of the correlator in \eqref{jjo3oddcft}. In the flat space limit we get :
\be\begin{split}\label{jjom1flat}
	\lim_{E \rightarrow 0}\ev{JJO_3}_{\text{odd}}&= \frac{k_3 }{E^2}\mathcal A_{\phi F\widetilde F}
\end{split}\ee
where $\mathcal A_{\phi F\widetilde F}$ is given in \eqref{Aphis}.


The corresponding parity-even correlator is given by \cite{Bzowski:2013sza,Bzowski:2018fql,Farrow:2018yni} :
\be\begin{split}\label{jjom}
\ev{JJO_3}_{\text{even}}&= \frac{(E+k_3) }{E^2}\mathcal M_{\phi F^2}
\end{split}\ee
This matches the expression for the homogeneous part of the correlator in \eqref{jjo3evencft} if we use the expression for $\mathcal M_{\phi F^2}$ in \eqref{fsqandfftilda}.
In the flat space limit the correlator takes the form :
\be\begin{split}\label{jjom1flat}
	\lim_{E \rightarrow 0}\ev{JJO_3}_{\text{even}}&= \frac{k_3 }{E^2}\mathcal A_{\phi F^2}
\end{split}\ee
where $\mathcal A_{\phi F^2}$ is given in \eqref{Aphis}.


\subsubsection{$\langle TTO_3\rangle$}
Let us now consider the parity-odd part of the correlator $\langle TTO_3\rangle$.
The only contribution to $\langle TTO_3\rangle_\text{odd}$ comes from the term :
\begin{align}
\int \phi~ W_{\alpha\beta\mu\nu}W^{\alpha\beta}_{~~\rho\sigma}~dx^{\mu} \wedge dx^{\nu} \wedge dx^{\rho} \wedge dx^{\sigma}
\end{align}
The corresponding $dS_4$ correlator is given by :
\begin{align}
\label{tto3cft}
\langle TTO_3\rangle_{\text{odd}}
&= k_1k_2\text{Im}\left[\int_{-\infty}^0 d\eta~ \eta^2 (1-ik_3 \eta)e^{i\eta E}\right]\mathcal{M}_{\phi W\widetilde{W}}\nonumber\\[5pt]
& = \frac{k_1k_2(E+3k_3)}{E^4}\mathcal{M}_{\phi W\widetilde{W}}
\end{align}
We can easily check that with the expression for $\mathcal{M}_{\phi W\widetilde{W}}$ in \eqref{mphiWWt}, this matches the homogeneous part of the correlator in \eqref{tto3oddcft}. In the flat space limit we get :
\begin{align}
\label{tto3cftflat}
\lim_{E \rightarrow 0}\langle TTO_3\rangle_{\text{odd}} = \frac{3k_1k_2k_3}{E^4}\mathcal{A}_{\phi W\widetilde{W}}
\end{align}
where $\mathcal A_{\phi W\widetilde{W}}$ is given in \eqref{AphiWWt}.

The parity-even part of the corresponding correlator is given by  \cite{Bzowski:2013sza,Bzowski:2018fql,Farrow:2018yni} :
\begin{align}
\ev{TTO}_{\text{even}}&= k_1 k_2\frac{E+3k_3}{E^4}\mathcal M_{\phi W^2}
\end{align}
which matches \eqref{tto3evencft} if we use the expression for $\mathcal M_{\phi W^2}$ given in \eqref{mphiWsquared}.
In the flat space limit we get :
\begin{align}
\lim_{E \rightarrow 0}\ev{TTO}_{\text{even}}&= \frac{3k_1k_2k_3}{E^4}\mathcal A_{\phi W^2}
\end{align}
where $\mathcal A_{\phi W^2}$ is given in \eqref{aphiw2}.

\subsubsection{$\langle JJJ\rangle$ }
Let us now compute the odd part of $\langle JJJ\rangle$. The contribution to $\langle JJJ\rangle_\text{odd}$ comes from the following terms :

\begin{align}
\int F_{\mu\tau}F_{\nu}^{~\tau}F_{\rho\sigma}~ dx^{\mu}\wedge dx^{\nu}\wedge dx^{\rho} \wedge dx^{\sigma}, \quad \int F_{\mu\nu}F_{\rho\sigma}~ dx^{\mu}\wedge dx^{\nu}\wedge dx^{\rho} \wedge dx^{\sigma}
\end{align}
The parts of the tree-level correlator corresponding to these interactions are given by :
\begin{align}
\mathcal{C}_{F^2\widetilde{F}} &= \text{Im}\left[\int_{-\infty}^0 d\eta~ \eta^2\,e^{i\eta E}\right]\mathcal{M}_{F^2\widetilde{F}}\nonumber\\[5pt]
\mathcal{C}_{F\widetilde{F}} &= \text{Im}\left[\int_{-\infty}^0 d\eta~ e^{i\eta E}\right]\mathcal{M}_{F\widetilde{F}}
\end{align}
Combining the two, we get the total correlator to be :
\begin{align}\label{JJJod1}
\langle JJJ\rangle_{\text{odd}}&=c_{JJJ}^{\text{odd}}\mathcal{C}_{F^2\widetilde{F}}+c_{JJ}^{\text{odd}}~\mathcal{C}_{F\widetilde{F}}\nonumber\\[5 pt]
&=c_{JJJ}^{\text{odd}}\frac{\mathcal{M}_{F^2\widetilde{F}}}{E^3}+c_{JJ}^{\text{odd}}~\frac{\mathcal{M}_{F\widetilde{F}}}{E}.
\end{align}
One can check using the explicit expression for $\mathcal{M}_{F^2\widetilde{F}}$ in \eqref{mF3F2Ft} that the first term in the above equation corresponds to the homogeneous term in \eqref{jjjoddcft}. Similarly, using the explicit expression for $\mathcal{M}_{F\widetilde{F}}$ in \eqref{mF2andmFFt} one can see that the second term in the above equation corresponds to the non-homogeneous term in \eqref{jjjoddcft}.
As noted earlier the term proportional to $c_{JJ}^{\text{odd}}$ in \eqref{JJJod1} is a contact term. In the flat space limit we obtain :
\begin{align}
\lim_{E\rightarrow 0}\langle JJJ\rangle_{\text{odd}}=c_{JJJ}^{\text{odd}}\frac{\mathcal{A}_{F^2\widetilde{F}}}{E^3}+c_{JJ}^{\text{odd}}~\frac{\mathcal{A}_{F\widetilde{F}}}{E}=c_{JJJ}^{\text{odd}}\frac{\mathcal{A}_{F^2\widetilde{F}}}{E^3}
\end{align}
where $\mathcal{A}_{F\widetilde{F}}=0$ as shown in \eqref{gluonamps1} and  the expression for $\mathcal A_{F^2\widetilde F}$ is as in \eqref{gluonamps}.
The even part of the correlator is given by \cite{Bzowski:2013sza,Bzowski:2017poo,Farrow:2018yni} :
\begin{align}\label{jjeva}
\ev{JJJ}_{\text{even}}
&=\frac{c_{JJJ}^{\text{even}}}{E^3}\mathcal M_{F^3}- \frac{2c_{JJ}^{\text{even}}}{E}\mathcal M_{YM}.
\end{align}
From the explicit expression for $\mathcal{M}_{F^3}$ in \eqref{mF3F2Ft1} we see that the first term in the above equation corresponds to the homogeneous term in \eqref{jjjevencft}. Using the expression for $\mathcal{M}_{YM}$ in \eqref{mF2andmFFt} one can see that the second term in the above equation corresponds to the non-homogeneous term in \eqref{jjjevencft}.
In the flat space limit we obtain :
\begin{align}
\lim_{E\rightarrow 0}\langle JJJ\rangle_{\text{even}}=\frac{c_{JJJ}^{\text{even}}}{E^3}\mathcal A_{F^3}- \frac{2c_{JJ}^{\text{even}}}{E}\mathcal A_{YM}
\end{align}
where the expression for $\mathcal A_{F^3}$ and $\mathcal A_{YM}$ are as in \eqref{gluonamps} and \eqref{gluonamps1} respectively.

\subsubsection{$\langle TTT\rangle$}
The contribution to $\langle TTT\rangle_\text{odd}$ comes from the following term in the action :
\begin{align}
\int W_{\rho\sigma\alpha\beta}W^{\sigma\gamma}_{~~\mu\tau}W_{\gamma~~~\nu}^{~\rho\tau}dx^{\mu}\wedge dx^{\nu} \wedge dx^{\alpha} \wedge dx^{\beta}
\end{align}
The tree-level $dS_4$ correlator corresponding to this term is given by :
\begin{align}
\langle TTT\rangle_\text{odd}
& =c_{TTT}^{\text{odd}} ~ k_1\,k_2\,k_3\,\text{Im}\left[\int_{-\infty}^0 d\eta\,\eta^5 e^{i\eta E}\right]\mathcal{M}_{W^2\widetilde{W}}\nonumber\\
&=c_{TTT}^{\text{odd}} ~ \frac{k_1\,k_2\,k_3\,}{E^6}\mathcal{M}_{W^2\widetilde{W}}
\end{align}
Using the result for $\mathcal{M}_{W^2\widetilde{W}}$ given in \eqref{A1}, one can check that this matches the homogeneous part of the correlator in \eqref{tto3oddcft}. This contribution was also calculated in \cite{Maldacena:2011nz}. In the flat space limit one obtains :
\begin{align}
\lim_{E\rightarrow 0}\langle TTT\rangle_\text{odd}=c_{TTT}^{\text{odd}} ~ \frac{k_1\,k_2\,k_3\,}{E^6}\mathcal{A}_{W^2\widetilde{W}}
\end{align}
where expression for $\mathcal{A}_{W^2\widetilde{W}}$ is given in \eqref{A2}.
The parity-even contribution to the correlator is given by \cite{Bzowski:2013sza,Bzowski:2017poo,Farrow:2018yni} :
\begin{align}\label{tttevena}
\ev{TTT}_{\text{even}}&=\frac{c_{TTT}^{\text{even}}c_{123}^2}{E^6}\mathcal M_{W^3}
+2c_{TT}^{\text{even}}\left(\frac{c_{123}}{E^2}+\frac{b_{123}}{E}-E\right)\mathcal M_{EG}
\end{align}
which upon using  $\mathcal M_{W^3}$  given in \eqref{MWWW} and  $\mathcal M_{EH}$ given in \eqref{EHM} produces \eqref{tttevencft} upto contact term\footnote{This expression matches upto a number with the expression for the stress tensor correlator that appears in equation $4.12$ of \cite{Farrow:2018yni} once we identity $$\mathcal{M}_{W^3} = -\frac{c_{123}E}{16J^2}\mathcal{A}_{F^3}\mathcal{M}_{F^3}.$$ See Appendix \ref{mf3mw3appendix} for details.}.

In the flat space limit we obtain
\begin{align}\label{tttevenaA}
\lim_{E\rightarrow 0} \ev{TTT}_{\text{even}}&=\frac{c_{TTT}^{\text{even}}c_{123}^2}{E^6}\mathcal A_{W^3}
+2c_{TT}^{\text{even}}\frac{c_{123}}{E^2}\mathcal A_{EG}
\end{align}
where $\mathcal A_{W^3}$ and $\mathcal A_{EG}$ are given in \eqref{A2} and \eqref{EHA}.

\subsection{Double copy structure of parity-violating amplitudes}
\label{DCamps}
In this section, we see how the double copy structure arises directly at the level of 3-point scattering amplitude when parity-violating terms are taken into account. The 3-point amplitudes involved may be written down directly on symmetry grounds, or computed from the action given above. The double copy relations of amplitudes also follows from that for CFT correlators that we saw previously, but here we show that it can be demonstrated directly in 4d flat space .

\subsection*{Scalar-graviton-graviton}
From the momentum space expressions for $\mathcal{M}_{\phi F^2}$ and $\mathcal{M}_{\phi F\widetilde{F}}$ in \eqref{fsqandfftilda} and the expression for $\mathcal{M}_{\phi W\widetilde{W}}$  and $\mathcal{M}_{\phi W^2}$ in \eqref{mphiWWt} and in \eqref{mphiWsquared} we obtain :
\begin{align} \label{DC1}
\mathcal{M}_{\phi W\widetilde{W}} = \mathcal{M}_{\phi F^2}\mathcal{M}_{\phi F\widetilde{F}},~~\mathcal{M}_{\phi W^2} &=  (\mathcal{M}_{\phi F^2})^2 = \mathcal{M}_{\phi F\widetilde{F}}\mathcal{M}_{\phi F\widetilde{F}}\nonumber\\
\mathcal{M}_{\phi W^2}+\mathcal{M}_{\phi W\widetilde{W}}&=(\mathcal M_{\phi F^2}+\mathcal{M}_{\phi F\widetilde{F}})^2
\end{align}
In the flat space limit when $E\rightarrow 0$ one has :
\begin{align}\label{DC1A}
\mathcal{A}_{\phi W\widetilde{W}} = \mathcal{A}_{\phi F^2}\mathcal{A}_{\phi F\widetilde{F}},~~\mathcal{M}_{\phi W^2} &=  (\mathcal{A}_{\phi F^2})^2 = \mathcal{A}_{\phi F\widetilde{F}}\mathcal{M}_{\phi F\widetilde{F}}\nonumber\\ 
\mathcal{A}_{\phi W^2}+\mathcal{A}_{\phi W\widetilde{W}}&=(\mathcal A_{\phi F^2}+\mathcal{A}_{\phi F\widetilde{F}})^2
\end{align}
%

\subsection*{Graviton-graviton-graviton}
From the momentum space expressions for $\mathcal{M}_{F^3}$ and $\mathcal{M}_{F^2\widetilde{F}}$ in \eqref{mF3F2Ft1} and \eqref{mF3F2Ft} and the expression for $\mathcal{M}_{W^2\widetilde{W}}$ in \eqref{A1} one can verify that : 
%
\begin{align}
\mathcal{M}_{W^2\widetilde{W}} = \mathcal{M}_{F^3}\mathcal{M}_{F^2\widetilde{F}}
\end{align}
In the $E\rightarrow 0$ limit one has :
\begin{align}
\mathcal{A}_{W^2\widetilde{W}} = 2\mathcal{A}_{F^3}\mathcal{A}_{F^2\widetilde{F}}
\end{align}
%
%
We also have the following relation : 
\begin{align}\label{mf3sq}
\mathcal{M}_{W^3} =\mathcal{M}^2_{F^3}
\end{align}
The details of how one obtains this relation in momentum space are given in Appendix \eqref{mf3mw3appendix}. Thus we notice that the following double copy structure
\begin{align}
\mathcal{A}_{W^3} =8\mathcal{A}^2_{F^3}
\end{align}
extends beyond the $E\rightarrow 0$ limit. We also notice using the momentum space expressions in \eqref{mF3F2Ft}  the following :
\begin{align}\label{MFMFt}
(\mathcal{M}_{F^3})^2 = (\mathcal{M}_{F^2\widetilde{F}})^2
\end{align}
Similarly one has the following double copy relation :
\be
\label{DC33}
\mathcal{M}_{W^3}  +  \mathcal{M}_{W^2\widetilde{W}}=(\mathcal{M}_{F^3}+ \mathcal{M}_{F^2\widetilde{F}})^2
\ee

\section{Discussion}
\label{discussion}
In this paper we established various double copy relations for parity-violating $CFT_3$ momentum space 3-point correlators. The double copy structure is a very special property of CFT correlators in momentum space, the analogue of which does not exist in position space. To understand this structure, we divided the momentum space CFT correlation function two parts, which we called homogeneous and non-homogeneous pieces.  It was crucial for our analysis that the homogeneous part consist of two pieces of conformally invariant structures, namely one parity-even and one parity-odd structure whereas the non-homogeneous part has only one parity-even conformally invariant piece - all other contributions are contact terms. Squaring the homogeneous piece could in principle generate three structures. However, interestingly it turns out that squaring the parity-odd and even part produces exactly the same structure, whereas the cross-term which is generated by multiplying the parity-odd and parity-even part gives rise to the needed parity-odd structure. 


This paper leaves various interesting directions to be followed upon. For example, for a general correlator of conserved currents of the form $\langle J_{s_1}J_{s_2}J_{s_3}\rangle$, we need to understand its structure better to be able make any detailed statement about its double copy. For instance, if we want to understand the parity-odd contribution we need to understand analogue of triangle inequality \cite{Giombi:2011rz}. We looked only at marginal scalars, but it is of interest to find out if the double copy structure shows up in $\ev{J_s J_s O_{\Delta}}$ if $O$ is a general scalar operator (that is, $\Delta$ is arbitrary).

It is also of interest to see whether the double copy relations continue to hold for higher point functions \cite{Armstrong:2020woi}. Establishing double copy relations for 4-point functions is a significantly harder problem. In this case it would be interesting to see if the double copy structure is visible even in some tractable limit or if one can infer the existence of a double copy structure for specific conformal blocks or Polyakov blocks. In this paper we have focused mainly on conformally invariant structures but it would be interesting to see what kind of constraints one needs to impose on OPE coefficients to get double copy relations.

Another interesting direction to study is a momentum space analogue of the analysis of \cite{Giombi:2011rz}.  In \cite{Giombi:2011rz} conformally invariant parity-even and parity-odd structures were identified in position space. These could then be appropriately composed to get conformally invariant parity-even and parity-odd correlators in position space. The double copy structure of the momentum space CFT correlators that we investigated in this paper already hints towards the existence of such structures in momentum space. See Appendix \ref{higherspin-correlator} for some details. We leave such questions for a future investigation.

\section*{Acknowledgments}
The work of SJ and RRJ is supported by the Ramanujan Fellowship. AM would like to acknowledge the support of CSIR-UGC
(JRF) fellowship (09/936(0212)/2019-EMR-I). The work of AS is supported by the KVPY scholarship.
We acknowledge our debt to the people of India for their steady support of research in basic sciences.
\appendix

\section{Expressions in spinor-helicity variables}
\label{SHvariablesapp}
In this section we write down various amplitudes in spinor helicity variables. Establishing a double copy at the level of spinor helicity variables is easy and obvious.
\subsubsection*{Gluon-gluon-scalar amplitudes}
In spinor-helicity variables, the non-zero components dS amplitude take the following form 
\begin{align}
\mathcal{M}_{\phi F^2}^{0--} &= 2\langle 12 \rangle^2,~~~
\mathcal{M}_{\phi F \widetilde{F}}^{0--}= 2i\langle 12 \rangle^2\label{mphiFFt}.
\end{align}
Also by complex conjugation $++$ helicity results also exist. However here as well as below we don't write them explicity. 
In spinor-helicity variables flat space amplitudes the take the form 
\begin{align}\label{mphiFFt1}
\mathcal{A}_{\phi F^2}^{0--} &= 2\langle 12 \rangle^2,~~~
\mathcal{A}_{\phi F \widetilde{F}}^{0--} = 2i\langle 12 \rangle^2
\end{align}
where we have written down answer for negative helicity component. 
Although \eqref{mphiFFt} and \eqref{mphiFFt1} might look similar, later is obtained from former in the $E\rightarrow 0$ limit.
\subsubsection*{Graviton-graviton-scalar amplitudes}
In spinor-helicity variables, the non-zero components of the $dS_4$ amplitude take the following form :
\begin{align}
\mathcal{M}^{0--}_{\phi W^2}= \langle12\rangle^4~~~
\mathcal{M}^{0--}_{\phi W\widetilde{W}}= 4i\langle12\rangle^4\label{mphiwwt}.
\end{align}
In spinor-helicity variables flat space amplitudes the take the form 
\begin{align}\label{mphiwwt1}
\mathcal{A}^{0--}_{\phi W^2}&= \langle12\rangle^4,~~~
\mathcal{A}^{0--}_{\phi W\widetilde{W}}= 4i\langle12\rangle^4
\end{align}
where we have written down answer for negative helicity component. 
Although \eqref{mphiwwt} and \eqref{mphiwwt1} might look similar, the latter is obtained from the former in the $E\rightarrow 0$ limit.

\subsubsection*{Double copy}
Comparing \eqref{mphiFFt} and \eqref{mphiFFt1} with \eqref{mphiwwt} and \eqref{mphiwwt1} we immediately see the double copy structure between gauge and gravity answers shown in \eqref{DC1} and \eqref{DC1A}.
\subsubsection*{Gluon-gluon-gluon amplitudes}
In spinor-helicity variables, the non-zero components of $dS$ amplitudes  are given by 
\begin{align}\label{mfs}
\mathcal{M}_{F^2}^{---} &= \frac{E}{k_1\,k_2\,k_3} \langle 12 \rangle \langle 23 \rangle \langle 31 \rangle,~~~~
\mathcal{M}_{F^2}^{--+} = \frac{(E-2k_3)}{k_1\,k_2\,k_3}\langle 12 \rangle \langle 2\bar{3} \rangle \langle \bar{3}1 \rangle\\[5 pt]
\mathcal{M}^{---}_{F\widetilde{F}} &= i\,\frac{E}{k_1\,k_2\,k_3}\,\langle 12 \rangle \langle 23 \rangle \langle 31 \label{mFFt}\rangle,~~
\mathcal{M}^{--+}_{F\widetilde{F}} = i\,\frac{E}{k_1\,k_2\,k_3}\,\langle 12 \rangle \langle 2\bar{3} \rangle \langle \bar{3}1 \rangle\\[5 pt]
\mathcal{M}_{F^3}^{---} &= \langle 12 \rangle \langle 23 \rangle \langle 31 \rangle,~~~~~~~
~~~~~~~~\mathcal{M}_{F^2 \widetilde{F}}^{---} = i\langle 12 \rangle \langle 23 \rangle \langle 31 \rangle
\end{align}

In the flat space limit we get
\begin{align}\label{mas}
\mathcal{A}_{F^2}^{---} &= 0,\,\,\,\mathcal{A}_{F^2}^{--+} = -2k_3\langle 12 \rangle \langle 2\bar{3} \rangle \langle \bar{3}1 \rangle\\[5pt]
\mathcal{A}^{---}_{F\widetilde{F}} &=0, \,\,\,\mathcal{A}^{--+}_{F\widetilde{F}} = 0 \nonumber\\
\mathcal{A}_{F^3}^{---} &= \langle 12 \rangle \langle 23 \rangle \langle 31 \rangle ,\,\,\,\, \mathcal{A}_{F^2 \widetilde{F}}^{---} = i\langle 12 \rangle \langle 23 \rangle \langle 31 \rangle
\end{align}

\subsubsection*{Graviton-graviton-graviton amplitudes}
In spinor-helicity variables, the non-zero components of $dS$ amplitudes  are given by 
\begin{align}\label{mwsp}
&\mathcal{M}^{---}_{W^3} = \langle 12\rangle^2\langle 23\rangle^2\langle 31\rangle^2,~~~
\mathcal{M}^{---}_{W^2\widetilde{W}} = i\langle 12\rangle^2\langle 23\rangle^2\langle 31\rangle^2
\end{align}
In the flat space limit we get
\begin{align}\label{masp}
&\mathcal{A}^{---}_{W^3} = \langle 12\rangle^2\langle 23\rangle^2\langle 31\rangle^2,~~~~
\mathcal{A}^{---}_{W^2\widetilde{W}} = i\langle 12\rangle^2\langle 23\rangle^2\langle 31\rangle^2
\end{align}
\subsubsection*{Double copy}
Comparing \eqref{mfs} and \eqref{mas} with \eqref{mwsp} and \eqref{masp} we immediately see the double copy structure between gauge and gravity answers shown in \eqref{DC33}.


\section{$\langle TJJ\rangle$}
\label{tjjapp}
 We calculate $\langle TJJ\rangle$ using gravity techniques for completeness. This result is not used in the main text.
\subsection{Mixed Gauge-Graviton amplitudes}
\subsection*{Gluon graviton interaction}
Let us  consider interactions between gluons and gravitons which contribute to mixed CFT correlators of the spin-one conserved current $J$ and the stress-tensor $T$. The interactions we consider are :
\begin{align}
S^I = \int d^4x \sqrt{g} F^2 + S^I_{\text{even}}+S^I_\text{odd}\label{IA}
\end{align}
where
\begin{align}
&S^I_{\text{even}} = \gamma\int d^4x\sqrt{g} ~W_{\mu\nu\rho\sigma}F^{\mu\nu}F^{\rho\sigma}\label{sieven}\\
&S^I_\text{odd} = \widetilde{\gamma}\int W_{\mu\nu\rho\sigma}F^{\mu\nu}F_{\alpha\beta} ~dx^{\rho}\wedge dx^{\sigma} \wedge dx^{\alpha} \wedge dx^{\beta}\label{siodd}
\end{align}
are the parity-preserving and parity-violating actions, respectively. 
Tree level scattering amplitude for parity-even case is given by
\begin{align}
\mathcal{A}_{F^2} &= -(z_1\cdot z_2)(z_1\cdot k_3)(z_3\cdot k_2)-(z_1\cdot k_2)(z_1\cdot z_3)(z_2\cdot k_3)+(z_1\cdot k_2)(z_2\cdot z_3)(z_1\cdot k_3)\nonumber\\[5pt]
\mathcal{A}_{WF^2} &= 4({z}_1\cdot{k}_2)({z}_2\cdot{k}_1) ({z}_1\cdot{k}_3)({z}_3\cdot{k}_1)\nonumber 
\end{align}

Let us now compute the odd part of $\langle TJJ\rangle$. The contribution to $\langle TJJ\rangle_\text{odd}$ comes from the following term in the action :
\begin{align}
\int W_{\mu\nu\rho\sigma}F^{\mu\nu}F_{\alpha\beta} ~dx^{\rho}\wedge dx^{\sigma} \wedge dx^{\alpha} \wedge dx^{\beta}
\end{align}
This gives parity 
\begin{align}
\mathcal{M}_{WF\widetilde{F}} &= \left[2({z}_1\cdot{k}_2)({z}_2\cdot{k}_1)+2(k_1\cdot k_2)({z}_1\cdot{z}_2)\right]\epsilon(z_1k_1z_3k_3)\nonumber\\[5pt]
&=\left[2(\vec{z}_1\cdot\vec{k}_2)(\vec{z}_2\cdot\vec{k}_1)+E(E-2k_3)\vec{z}_1\cdot\vec{z}_2\right]\epsilon(z_1k_1z_3k_3)
\end{align}
The flat space amplitudes  are obtained by taking $E \rightarrow 0$ :
\begin{align}
\mathcal{A}_{WF\widetilde{F}} =2({z}_1\cdot{k}_2)({z}_2\cdot{k}_1)\epsilon(z_1k_1z_3k_3)
\end{align}
The non-zero spinor-helicity components of the above ds amplitude  are given by
\begin{align}
\label{CBG}
&\mathcal{M}^{--+}_{F^2} = \frac{\langle 12\rangle^2\langle1\bar{3}\rangle^2}{2 k^2_1},~~
\mathcal{M}^{---}_{WF^2} = 4\langle12\rangle^2\langle13\rangle^2,~~
\mathcal{M}^{---}_{WF\widetilde{F}} = 4\langle12\rangle^2\langle13\rangle^2.
\end{align}
In the spinor-helicity language the non-zero components of flat space amplitude  are given by 
\begin{align}
&\mathcal{A}^{--+}_{F^2} = \frac{\langle 12\rangle^2\langle1\bar{3}\rangle^2}{2 k^2_1},~~
\mathcal{A}^{---}_{WF^2} = 4\langle12\rangle^2\langle13\rangle^2,~~
\mathcal{A}^{---}_{WF\widetilde{F}} = 4\langle12\rangle^2\langle13\rangle^2.
\end{align}
Now the parity-odd contribution to correlator $\langle TJJ\rangle$ is given by
\begin{align}\label{JJT1}
\langle TJJ\rangle_{\text{odd}}&= k_1\text{Im}\left[\int_{-\infty}^0 d\eta ~\eta^3 e^{i\eta E} \right] \mathcal{M}_{WF\widetilde{F}} \nonumber\\[5pt]
&= \frac{k_1}{E^4}\mathcal{M}_{WF\widetilde{F}}
\end{align}
One can check that the correlation function given in \eqref{JJT1}  satisfies appropriate conformal Ward identity.

\section{Proof of some double copy relations}
\label{DCapp}
\subsection{$\mathcal{M}_{W^3} \propto (\mathcal{M}_{F^3})^2$}\label{mf3mw3appendix}
Here, we briefly show how \eqref{mf3sq} can be derived in momentum space. 
$\mathcal{M}_{F^3}^2$ is given by
\begin{align}\label{mf32}
\mathcal{M}_{F^3}^2 &= 4(\vec{k}_2 \cdot \vec{z}_1)^2(\vec{k}_1 \cdot \vec{z}_3)^2(\vec{k}_3 \cdot \vec{z}_2)^2 +4E\left((\vec{k}_2 \cdot \vec{z}_1)^2 (\vec{k}_1 \cdot \vec{z}_3)(\vec{k}_3 \cdot \vec{z}_2)(\vec{z}_2 \cdot \vec{z}_3)+ \text{cyclic perm.}\right)\nonumber \\[5 pt]
&+E^2\left(2k_1 k_2 (\vec{k}_2  \cdot \vec{z}_1)(\vec{k}_3 \cdot \vec{z}_2)(\vec{z}_1 \cdot \vec{z}_3)(\vec{z}_2 \cdot \vec{z}_3)+k_1^2 (\vec{k}_2 \cdot \vec{z}_1)^2 (\vec{z}_2 \cdot \vec{z}_3)^2 +\text{cyclic perm.}\right)
\end{align} 
$\mathcal{M}_{W^3}$ is given by
\begin{align}\label{mw3}
\begin{split}
\mathcal{M}_{W^3}=& A_{1}\left(k_{1}, k_{2}, k_{3}\right)(\vec{k}_2 \cdot \vec{z}_1)^2 (\vec{k}_3 \cdot \vec{z}_2)^2(\vec{k}_1 \cdot \vec{z}_3)^2 \\
&+\left(A_{2}\left(k_{1}, k_{2}, k_{3}\right) (\vec{z}_1 \cdot \vec{z}_2) (\vec{k}_2 \cdot \vec{z}_1))(\vec{k}_3 \cdot \vec{z}_2)\left(\vec{k}_1 \cdot \vec{z}_3\right)^{2}+\text{cyclic perm.}\right) \\
&+\left(A_{3}\left(k_{1}, k_{2}, k_{3}\right)\left(\vec{z}_1 \cdot \vec{z}_2\right)^{2}\left(\vec{k}_1 \cdot \vec{z}_3\right)^{2}+\text{cyclic perm.}\right) \\
&+\left(A_{4}\left(k_{1}, k_{2}, k_{3}\right) (\vec{z}_1 \cdot \vec{z}_3) (\vec{z}_2 \cdot \vec{z}_3) (\vec{k}_2 \cdot \vec{z}_1) (\vec{k}_3 \cdot \vec{z}_2)+\text{cyclic perm.}\right) \\
&+A_{5}\left(k_{1}, k_{2}, k_{3}\right) (\vec{z}_1 \cdot \vec{z}_2)(\vec{z}_2 \cdot \vec{z}_3)(\vec{z}_1 \cdot \vec{z}_3)
\end{split}
\end{align}
where \cite{Farrow:2018yni} $$A_{1}=8, \quad A_{2}=4 E\left(2 k_{3}-E\right), \quad A_{3}=0, \quad A_{4}=2 E^{2}\left(E-2 k_{1}\right)\left(E-2 k_{1}\right), \quad A_{5}=-J^{2} E^{2}$$
Using the following two degeneracies \cite{Bzowski:2013sza} to set $A_4$ and $A_5$ to zero in \eqref{mw3}, it was shown that $\mathcal{M}_{W^3}$ can be related to $\mathcal{A}_{F^3} \mathcal{M}_{F^3}$. 
\begin{align}
&(\vec{k}_2 \cdot \vec{z}_1)^2 (\vec{k}_3 \cdot \vec{z}_2)^2 +(k_3^2-k_1^2-k_2^2)(\vec{k}_2 \cdot \vec{z}_1)(\vec{k}_3 \cdot \vec{z}_2)(\vec{z}_1 \cdot \vec{z}_2)-\frac{J^2}{4}(\vec{z}_1 \cdot \vec{z}_2)^2 =0 \label{degen1}\\[5 pt]
&(\vec{k}_2 \cdot \vec{z}_1)(\vec{k}_3 \cdot \vec{z}_2)(\vec{k}_1 \cdot \vec{z}_3)^2 + k_3^2 (\vec{z}_1 \cdot \vec{z}_2) (\vec{k}_1 \cdot \vec{z}_3)^2 +\frac{1}{2}(k_1^2-k_2^2+k_3^2)(\vec{z}_2 \cdot \vec{z}_3)(\vec{k}_2 \cdot \vec{z}_1)(\vec{k}_1 \cdot \vec{z}_3)\nonumber \\[5 pt]
&+\frac{1}{2}(k_2^2-k_1^2+k_3^2)(\vec{z}_1 \cdot \vec{z}_3)(\vec{k}_3 \cdot \vec{z}_2)(\vec{k}_1 \cdot \vec{z}_3)+\frac{J^2}{4}(\vec{z}_1 \cdot \vec{z}_3)(\vec{z}_2 \cdot \vec{z}_3) =0\label{degen2}
\end{align}
The relation is given by
\begin{align}
\mathcal{M}_{W^3} = -\frac{c_{123}E}{16J^2}\mathcal{A}_{F^3}\mathcal{M}_{F^3} 
\end{align}
where
\begin{align}
\mathcal{A}_{F^3}\mathcal{M}_{F^3} = 4(\vec{k}_2 \cdot \vec{z}_1)^2(\vec{k}_1 \cdot \vec{z}_3)^2(\vec{k}_3 \cdot \vec{z}_2)^2 +2E\left((\vec{k}_2 \cdot \vec{z}_1)^2 (\vec{k}_1 \cdot \vec{z}_3)(\vec{k}_3 \cdot \vec{z}_2)(\vec{z}_2 \cdot \vec{z}_3)k_1+ \text{cyclic perm.}\right)
\end{align}
Using the same two degeneracies on \eqref{mf32} to remove the term proportional to $E^2$, we obtain the following relation.
\begin{align}
-8\frac{c_{123}E}{J^2}\mathcal{A}_{F^3}\mathcal{M}_{F^3} = (\mathcal{M}_{F^3})^2 = \frac{1}{2}\mathcal{M}_{W^3}
\end{align}
Therefore, we see that 
\begin{align}
(\mathcal{M}_{F^3})^2 \propto \mathcal{M}_{W^3}
\end{align}
which says that the double copy holds beyond the flat space limit.

\subsection{$(\mathcal{M}_{\phi F^2})^2 \propto (\mathcal{M}_{\phi F\widetilde{F}})^2$}
 Some of the  equality relation in (\ref{DC1}) was established in \cite{Farrow:2018yni}. Here we establish the second part of (\ref{DC1}). Consider the identity
\begin{align}
\epsilon_{\mu\nu\rho\sigma}\epsilon^{\alpha\beta\gamma\delta} = 
\begin{vmatrix}
\delta^{\alpha}_{\mu} && \delta_{\mu}^{\beta} && \delta_{\mu}^{\gamma} && \delta_{\mu}^{\delta}\\
\delta^{\alpha}_{\nu} && \delta_{\nu}^{\beta} && \delta_{\nu}^{\gamma} && \delta_{\nu}^{\delta}\\
\delta^{\alpha}_{\rho} && \delta_{\rho}^{\beta} && \delta_{\rho}^{\gamma} && \delta_{\rho}^{\delta}\\
\delta^{\alpha}_{\sigma} && \delta_{\sigma}^{\beta} && \delta_{\sigma}^{\gamma} && \delta_{\sigma}^{\delta}\\
\end{vmatrix}\label{ei}
\end{align}
Since, we have
\begin{align}
\mathcal{M}_{\phi F\widetilde{F}} = \epsilon(z_1 k_1 z_2k_2)
\end{align}
Therefore, we may write
\begin{align}
(\mathcal{M}_{\phi F\widetilde{F}})^2 = 
\begin{vmatrix}
0 && 0 && \vec{z}_1. \vec{z}_2 && \vec{z}_1. \vec{k}_2\\
0 && 0 && \vec{z}_2.\vec{k}_1 &&k_1.k_2\\
\vec{z}_1.\vec{z}_2 && \vec{z}_2.\vec{k}_1 && 0 && 0\\
\vec{z}_1.\vec{k}_2 && k_1.k_2 && 0 && 0\\
\end{vmatrix}
= [(\vec{z}_1.\vec{z}_2) k_1.k_2-\vec{z}_1.\vec{k}_2 \vec{z}_2.\vec{k}_1]^2
\end{align}
Since, in 4D we have $k_1.k_2 = -\frac{E(E-2k_3)}{2}$, using \eqref{mphiWsquared}
\begin{align}
(\mathcal{M}_{\phi F\widetilde{F}})^2 \propto(\mathcal{M}_{\phi F^2})^2 
\end{align}

\subsection{$(\mathcal{M}_{F^3})^2 \propto (\mathcal{M}_{F^2\widetilde{F}})^2$}
Here we establish the relation (\ref{MFMFt}). Consider the identity (\ref{ei}). Since, we have
\begin{align}
\mathcal{M}_{F^2\widetilde{F}} = \epsilon^{\mu\nu\rho\sigma}{F_{(1)}}_{\mu}^{~\tau}F_{(2)\tau\nu}F_{(3)\rho\sigma} 
\end{align}
Therefore, we may write
\begin{align}
(\mathcal{M}_{F^2\widetilde{F}})^2 &= \epsilon^{\mu\nu\rho\sigma}{F_{(1)}}_{\mu}^{~\tau}{F_{(2)}}_{\tau\nu}{F_{(3)}}_{\rho\sigma}\epsilon^{\alpha\beta\gamma\delta}{F_{(1)}}_{\alpha}^{~\lambda}{F_{(2)}}_{\lambda\beta}{F_{(3)}}_{\gamma\delta}\notag\\&=
\begin{vmatrix}
0 && \delta_{\nu\alpha} && z_{3\alpha} && k_{3\alpha}\\
\delta_{\mu\beta} && 0 && z_{3\beta} && k_{3\beta}\\
z_{3\mu} && z_{3\nu} && 0 && 0\\
k_{3\mu} && k_{3\nu} && 0 && 0\\
\end{vmatrix}
{F_{(1)}}_{\mu}^{~\tau}{F_{(2)}}_{\tau\nu}{F_{(1)}}_{\alpha}^{~\lambda}{F_{(2)}}_{\lambda\beta}
\notag\\
&=({F_{(3)}}^{\mu\nu} {F_{(3)}}^{\alpha\beta}){F_{(1)}}_{\mu}^{~\tau}{F_{(2)}}_{\tau\nu}{F_{(1)}}_{\alpha}^{~\lambda}{F_{(2)}}_{\lambda\beta} = (\mathcal{M}_{F^3})^2
\end{align}

\section{Momentum space expression of higher spin correlators}\label{higherspin-correlator}
In this section we give the momentum space expression for the parity-even and parity-odd parts of higher spin correlators $\langle J_sJ_sO_3\rangle$ and $\langle J_sJ_sJ_s\rangle$ using relations in subsection \ref{hs1} and the the expressions for $\langle JJO_3\rangle$ and $\langle JJJ\rangle$ given in Section \ref{cftcorrelators}.
\begin{align}\label{spinsintermsofspin1}
\langle J_sJ_sO_3\rangle_{\text{even},\bf{h}}
&=(k_1 k_2)^{s-1}(E+(2s-1)k_3)\left[\frac{1}{E^2} \left\{2(\vec{z}_1\cdot \vec{k}_2)( \vec{z}_2\cdot \vec{k}_1) +E (E-2k_3)\vec{z}_1\cdot \vec{z}_2 \right\}\right]^s\cr
\langle J_sJ_sO_3\rangle_{\text{odd},\bf{h}}
&= (k_1 k_2)^{s-1}\frac{(E+(2s-1)k_3)}{E^{2s}}\left[k_2\,\epsilon^{ k_1 z_1 z_2}- k_1\,\epsilon^{k_2 z_1  z_2}\right]\cr
&\hspace{.5cm}\times \left[2(\vec{z}_1\cdot \vec{k}_2)( \vec{z}_2\cdot \vec{k}_1) +E (E-2k_3)\vec{z}_1\cdot \vec{z}_2 \right]^{s-1}\cr
\langle J_sJ_sJ_s\rangle_{\text{even},\bf{h}}
&=(k_1 k_2 k_3)^{s-1}\left[\frac{1}{E^3} \Big\{2\,(\vec{z}_1\cdot \vec{k}_2) \, (\vec{z}_2\cdot \vec{k}_3) \, (\vec{z}_3\cdot \vec{k}_1)+E \{k_3\, (\vec{z}_1\cdot \vec{z}_2) \, (\vec{z}_3\cdot \vec{k}_1)+ \text{cyclic}\}\Big\}\right]^s\cr
\langle J_sJ_sJ_s\rangle_{\text{odd},\bf{h}}
&=(k_1 k_2 k_3)^{s-1}\frac{1}{E^3}\left[\left\{(\vec{k}_1 \cdot \vec{z}_3)\left(\epsilon^{k_3 z_1 z_2}k_1-\epsilon^{k_1 z_1 z_2}k_3\right)+(\vec{k}_3 \cdot \vec{z}_2)\left(\epsilon^{k_1 z_1 z_3}k_2-\epsilon^{k_2 z_1 z_3}k_1\right)\right.\right.\nonumber\\[5 pt]
&\hspace{1.5cm}\left.\left.-(\vec{z}_2 \cdot \vec{z}_3)\epsilon^{k_1 k_2 z_1}E+\frac{k_1}{2} \epsilon^{z_1 z_2 z_3}E(E-2k_1)\right\}+\text{cyclic perm}\right]\cr
&\hspace{.5cm}\times\left[\frac{1}{E^3} \Big\{2\,(\vec{z}_1\cdot \vec{k}_2) \, (\vec{z}_2\cdot \vec{k}_3) \, (\vec{z}_3\cdot \vec{k}_1)+E \{k_3\, (\vec{z}_1\cdot \vec{z}_2) \, (\vec{z}_3\cdot \vec{k}_1)+ \text{cyclic}\}\Big\}\right]^{s-1}.\nonumber\\
\end{align}
One can use the Todorov operator \cite{Dobrev:1975ru} to strip off the polarization vectors from the expressions in \eqref{spinsintermsofspin1}. This operator is given by
\begin{align}
D_{z}^{i}=\left(\frac{1}{2}+\vec{z} \cdot \frac{\partial}{\partial \vec{z}}\right) \frac{\partial}{\partial z_{i}}-\frac{1}{2} z^{i} \frac{\partial^{2}}{\partial \vec{z} \cdot \partial \vec{z}}
\end{align}
Therefore, we have
\begin{align}
\langle J^{\mu_1 \cdots \mu_s}J^{\nu_1 \cdots \nu_s}O_3\rangle_{\text{even},\bf{h}}&=\prod_{i=1}^{s}{D_{z_1}^{\mu_i}D_{z_2}^{\nu_i}}\langle J_sJ_sO_3\rangle_{\text{even},\bf{h}} \notag\\
\langle J^{\mu_1 \cdots \mu_s}J^{\nu_1 \cdots \nu_s}O_3\rangle_{\text{odd},\bf{h}}&=\prod_{i=1}^{s}{D_{z_1}^{\mu_i}D_{z_2}^{\nu_i}}\langle J_sJ_sO_3\rangle_{\text{odd},\bf{h}} \notag\\
\langle J^{\mu_1 \cdots \mu_s}J^{\nu_1 \cdots \nu_s}J^{\rho_1 \cdots \rho_s}\rangle_{\text{even},\bf{h}} &= \prod_{i=1}^{s}{D_{z_1}^{\mu_i}D_{z_2}^{\nu_i}D_{z_3}^{\rho_i}}\langle J_sJ_sJ_s\rangle_{\text{even},\bf{h}} \notag \\
\langle J^{\mu_1 \cdots \mu_s}J^{\nu_1 \cdots \nu_s}J^{\rho_1 \cdots \rho_s}\rangle_{\text{odd},\bf{h}} &= \prod_{i=1}^{s}{D_{z_1}^{\mu_i}D_{z_2}^{\nu_i}D_{z_3}^{\rho_i}}\langle J_sJ_sJ_s\rangle_{\text{odd},\bf{h}} 
\end{align}

\section{Double copy relations in spinor-helicity notation}
In our discussion of double copy relations we have mostly focussed on the momentum dependent structures of the correlators. However, a strict doubly copy relation would in addition relate the OPE like coefficients in the correlators. In this section we make this clear using the spinor-helicity notation and the double copy structure of $\langle J_{2s}J_{2s}J_{2s}\rangle$.

We have the following non-zero spinor-helicity components of $\langle J_sJ_sJ_s\rangle$ :
\begin{align}
\langle J_{s}^-J_{s}^-J_{s}^-\rangle_{\bf{h}}=(c_{s,{\text{even}}}+i\,c_{s,{\text{odd}}})\frac{(k_1k_2k_3)^{s-1}}{E^{3s}}\langle 12\rangle^s\langle 23\rangle^s\langle 31\rangle^s\cr
\langle J_{s}^+J_{s}^+J_{s}^+\rangle_{\bf{h}}=(c_{s,{\text{even}}}-i\,c_{s,{\text{odd}}})\frac{(k_1k_2k_3)^{s-1}}{E^{3s}}\langle \bar 1\bar 2\rangle^s\langle \bar 2\bar 3\rangle^s\langle \bar 3\bar 1\rangle^s
\end{align}
The non-zero spinor-helicity components of $\langle JJJ\rangle$ are :
\begin{align}
\langle J^-J^-J^-\rangle_{\bf{h}}=(c_{1,{\text{even}}}+i\,c_{1,{\text{odd}}})\frac{1}{E^{3}}\langle 12\rangle\langle 23\rangle\langle 31\rangle\cr
\langle J^+J^+J^+\rangle_{\bf{h}}=(c_{1,{\text{even}}}-i\,c_{1,{\text{odd}}})\frac{1}{E^{3}}\langle \bar 1\bar 2\rangle\langle \bar 2\bar 3\rangle\langle \bar 3\bar 1\rangle
\end{align}
Demanding the $s$-copy relations of subsection \eqref{hs1} we get the following equations that constrain the coefficients $c_{s,{\text{even}}}$ and $c_{s,{\text{odd}}}$ in terms of $c_{1,{\text{even}}}$ and $c_{1,{\text{even}}}$ :
\begin{align}
c_{s,\text{even}}+i\,c_{s,\text{odd}}&=(c_{1,{\text{even}}}+i\,c_{1,{\text{odd}}})^s\cr
c_{s,\text{even}}-i\,c_{s,\text{odd}}&=(c_{1,{\text{even}}}-i\,c_{1,{\text{odd}}})^s
\end{align}
Specifically when $s=2$, for which the conserved current is the stress tensor we obtain :
\begin{align}
c_{TTT,\text{even}}&=c_{1,{\text{even}}}^2-c_{1,{\text{odd}}}^2\cr
c_{TTT,\text{odd}}&=2c_{1,{\text{even}}}c_{1,{\text{odd}}}
\end{align}

\bibliography{DC}

\providecommand{\href}[2]{#2}\begingroup\raggedright\begin{thebibliography}{10}

\bibitem{Gary:2009ae}
M.~Gary, S.~B. Giddings and J.~Penedones, \emph{{Local bulk S-matrix elements
  and CFT singularities}},
  \href{https://doi.org/10.1103/PhysRevD.80.085005}{\emph{Phys. Rev. D}
  {\bfseries 80} (2009) 085005}
  [\href{https://arxiv.org/abs/0903.4437}{{\ttfamily 0903.4437}}].

\bibitem{Gary:2009mi}
M.~Gary and S.~B. Giddings, \emph{{The Flat space S-matrix from the AdS/CFT
  correspondence?}},
  \href{https://doi.org/10.1103/PhysRevD.80.046008}{\emph{Phys. Rev. D}
  {\bfseries 80} (2009) 046008}
  [\href{https://arxiv.org/abs/0904.3544}{{\ttfamily 0904.3544}}].

\bibitem{Komatsu:2020sag}
S.~Komatsu, M.~F. Paulos, B.~C. Van~Rees and X.~Zhao, \emph{{Landau diagrams in
  AdS and S-matrices from conformal correlators}},
  \href{https://doi.org/10.1007/JHEP11(2020)046}{\emph{JHEP} {\bfseries 11}
  (2020) 046} [\href{https://arxiv.org/abs/2007.13745}{{\ttfamily
  2007.13745}}].

\bibitem{Penedones:2010ue}
J.~Penedones, \emph{{Writing CFT correlation functions as AdS scattering
  amplitudes}}, \href{https://doi.org/10.1007/JHEP03(2011)025}{\emph{JHEP}
  {\bfseries 03} (2011) 025} [\href{https://arxiv.org/abs/1011.1485}{{\ttfamily
  1011.1485}}].

\bibitem{Raju:2012zr}
S.~Raju, \emph{{New Recursion Relations and a Flat Space Limit for AdS/CFT
  Correlators}}, \href{https://doi.org/10.1103/PhysRevD.85.126009}{\emph{Phys.
  Rev. D} {\bfseries 85} (2012) 126009}
  [\href{https://arxiv.org/abs/1201.6449}{{\ttfamily 1201.6449}}].

\bibitem{Fitzpatrick:2011hu}
A.~L. Fitzpatrick and J.~Kaplan, \emph{{Analyticity and the Holographic
  S-Matrix}}, \href{https://doi.org/10.1007/JHEP10(2012)127}{\emph{JHEP}
  {\bfseries 10} (2012) 127} [\href{https://arxiv.org/abs/1111.6972}{{\ttfamily
  1111.6972}}].

\bibitem{Fitzpatrick:2011dm}
A.~L. Fitzpatrick and J.~Kaplan, \emph{{Unitarity and the Holographic
  S-Matrix}}, \href{https://doi.org/10.1007/JHEP10(2012)032}{\emph{JHEP}
  {\bfseries 10} (2012) 032} [\href{https://arxiv.org/abs/1112.4845}{{\ttfamily
  1112.4845}}].

\bibitem{Caron-Huot:2017vep}
S.~Caron-Huot, \emph{{Analyticity in Spin in Conformal Theories}},
  \href{https://doi.org/10.1007/JHEP09(2017)078}{\emph{JHEP} {\bfseries 09}
  (2017) 078} [\href{https://arxiv.org/abs/1703.00278}{{\ttfamily
  1703.00278}}].

\bibitem{Gillioz:2020mdd}
M.~Gillioz, M.~Meineri and J.~Penedones, \emph{{A scattering amplitude in
  Conformal Field Theory}},
  \href{https://doi.org/10.1007/JHEP11(2020)139}{\emph{JHEP} {\bfseries 11}
  (2020) 139} [\href{https://arxiv.org/abs/2003.07361}{{\ttfamily
  2003.07361}}].

\bibitem{Kawai:1985xq}
H.~Kawai, D.~C. Lewellen and S.~H.~H. Tye, \emph{{A Relation Between Tree
  Amplitudes of Closed and Open Strings}},
  \href{https://doi.org/10.1016/0550-3213(86)90362-7}{\emph{Nucl. Phys. B}
  {\bfseries 269} (1986) 1}.

\bibitem{Bern:2008qj}
Z.~Bern, J.~J.~M. Carrasco and H.~Johansson, \emph{{New Relations for
  Gauge-Theory Amplitudes}},
  \href{https://doi.org/10.1103/PhysRevD.78.085011}{\emph{Phys. Rev. D}
  {\bfseries 78} (2008) 085011}
  [\href{https://arxiv.org/abs/0805.3993}{{\ttfamily 0805.3993}}].

\bibitem{Bern:2010ue}
Z.~Bern, J.~J.~M. Carrasco and H.~Johansson, \emph{{Perturbative Quantum
  Gravity as a Double Copy of Gauge Theory}},
  \href{https://doi.org/10.1103/PhysRevLett.105.061602}{\emph{Phys. Rev. Lett.}
  {\bfseries 105} (2010) 061602}
  [\href{https://arxiv.org/abs/1004.0476}{{\ttfamily 1004.0476}}].

\bibitem{Broedel:2012rc}
J.~Broedel and L.~J. Dixon, \emph{{Color-kinematics duality and double-copy
  construction for amplitudes from higher-dimension operators}},
  \href{https://doi.org/10.1007/JHEP10(2012)091}{\emph{JHEP} {\bfseries 10}
  (2012) 091} [\href{https://arxiv.org/abs/1208.0876}{{\ttfamily 1208.0876}}].

\bibitem{Johansson:2017srf}
H.~Johansson and J.~Nohle, \emph{{Conformal Gravity from Gauge Theory}},
  \href{https://arxiv.org/abs/1707.02965}{{\ttfamily 1707.02965}}.

\bibitem{Johansson:2018ues}
H.~Johansson, G.~Mogull and F.~Teng, \emph{{Unraveling conformal gravity
  amplitudes}}, \href{https://doi.org/10.1007/JHEP09(2018)080}{\emph{JHEP}
  {\bfseries 09} (2018) 080}
  [\href{https://arxiv.org/abs/1806.05124}{{\ttfamily 1806.05124}}].

\bibitem{Bern:2013yya}
Z.~Bern, S.~Davies, T.~Dennen, Y.-t. Huang and J.~Nohle,
  \emph{{Color-Kinematics Duality for Pure Yang-Mills and Gravity at One and
  Two Loops}}, \href{https://doi.org/10.1103/PhysRevD.92.045041}{\emph{Phys.
  Rev. D} {\bfseries 92} (2015) 045041}
  [\href{https://arxiv.org/abs/1303.6605}{{\ttfamily 1303.6605}}].

\bibitem{He:2017spx}
S.~He, O.~Schlotterer and Y.~Zhang, \emph{{New BCJ representations for one-loop
  amplitudes in gauge theories and gravity}},
  \href{https://doi.org/10.1016/j.nuclphysb.2018.03.003}{\emph{Nucl. Phys. B}
  {\bfseries 930} (2018) 328}
  [\href{https://arxiv.org/abs/1706.00640}{{\ttfamily 1706.00640}}].

\bibitem{Bern:2019prr}
Z.~Bern, J.~J. Carrasco, M.~Chiodaroli, H.~Johansson and R.~Roiban, \emph{{The
  Duality Between Color and Kinematics and its Applications}},
  \href{https://arxiv.org/abs/1909.01358}{{\ttfamily 1909.01358}}.

\bibitem{Coriano:2013jba}
C.~Coriano, L.~Delle~Rose, E.~Mottola and M.~Serino, \emph{{Solving the
  Conformal Constraints for Scalar Operators in Momentum Space and the
  Evaluation of Feynman's Master Integrals}},
  \href{https://doi.org/10.1007/JHEP07(2013)011}{\emph{JHEP} {\bfseries 07}
  (2013) 011} [\href{https://arxiv.org/abs/1304.6944}{{\ttfamily 1304.6944}}].

\bibitem{Bzowski:2013sza}
A.~Bzowski, P.~McFadden and K.~Skenderis, \emph{{Implications of conformal
  invariance in momentum space}},
  \href{https://doi.org/10.1007/JHEP03(2014)111}{\emph{JHEP} {\bfseries 03}
  (2014) 111} [\href{https://arxiv.org/abs/1304.7760}{{\ttfamily 1304.7760}}].

\bibitem{Bonora:2015nqa}
L.~Bonora, A.~D. Pereira and B.~Lima~de Souza, \emph{{Regularization of
  energy-momentum tensor correlators and parity-odd terms}},
  \href{https://doi.org/10.1007/JHEP06(2015)024}{\emph{JHEP} {\bfseries 06}
  (2015) 024} [\href{https://arxiv.org/abs/1503.03326}{{\ttfamily
  1503.03326}}].

\bibitem{Bonora:2015odi}
L.~Bonora and B.~Lima~de Souza, \emph{{Pure contact term correlators in CFT}},
  {\emph{Bled Workshops Phys.} {\bfseries 16} (2015) 22}
  [\href{https://arxiv.org/abs/1511.06635}{{\ttfamily 1511.06635}}].

\bibitem{Bonora:2016ida}
L.~Bonora, M.~Cvitan, P.~Dominis~Prester, B.~Lima~de Souza and I.~Smoli\'c,
  \emph{{Massive fermion model in 3d and higher spin currents}},
  \href{https://doi.org/10.1007/JHEP05(2016)072}{\emph{JHEP} {\bfseries 05}
  (2016) 072} [\href{https://arxiv.org/abs/1602.07178}{{\ttfamily
  1602.07178}}].

\bibitem{sissathesis}
L.~de~Souza., \emph{{CFT’s, contact terms and anomalies}},
  \href{https://arxiv.org/abs/PhD Thesis}{{\ttfamily PhD Thesis}}.

\bibitem{Bzowski:2015pba}
A.~Bzowski, P.~McFadden and K.~Skenderis, \emph{{Scalar 3-point functions in
  CFT: renormalisation, beta functions and anomalies}},
  \href{https://doi.org/10.1007/JHEP03(2016)066}{\emph{JHEP} {\bfseries 03}
  (2016) 066} [\href{https://arxiv.org/abs/1510.08442}{{\ttfamily
  1510.08442}}].

\bibitem{Bzowski:2015yxv}
A.~Bzowski, P.~McFadden and K.~Skenderis, \emph{{Evaluation of conformal
  integrals}}, \href{https://doi.org/10.1007/JHEP02(2016)068}{\emph{JHEP}
  {\bfseries 02} (2016) 068}
  [\href{https://arxiv.org/abs/1511.02357}{{\ttfamily 1511.02357}}].

\bibitem{Bzowski:2017poo}
A.~Bzowski, P.~McFadden and K.~Skenderis, \emph{{Renormalised 3-point functions
  of stress tensors and conserved currents in CFT}},
  \href{https://doi.org/10.1007/JHEP11(2018)153}{\emph{JHEP} {\bfseries 11}
  (2018) 153} [\href{https://arxiv.org/abs/1711.09105}{{\ttfamily
  1711.09105}}].

\bibitem{Coriano:2018bbe}
C.~Corian\`o and M.~M. Maglio, \emph{{Exact Correlators from Conformal Ward
  Identities in Momentum Space and the Perturbative $TJJ$ Vertex}},
  \href{https://doi.org/10.1016/j.nuclphysb.2018.11.016}{\emph{Nucl. Phys. B}
  {\bfseries 938} (2019) 440}
  [\href{https://arxiv.org/abs/1802.07675}{{\ttfamily 1802.07675}}].

\bibitem{Bzowski:2018fql}
A.~Bzowski, P.~McFadden and K.~Skenderis, \emph{{Renormalised CFT 3-point
  functions of scalars, currents and stress tensors}},
  \href{https://doi.org/10.1007/JHEP11(2018)159}{\emph{JHEP} {\bfseries 11}
  (2018) 159} [\href{https://arxiv.org/abs/1805.12100}{{\ttfamily
  1805.12100}}].

\bibitem{Gillioz:2018mto}
M.~Gillioz, \emph{{Momentum-space conformal blocks on the light cone}},
  \href{https://doi.org/10.1007/JHEP10(2018)125}{\emph{JHEP} {\bfseries 10}
  (2018) 125} [\href{https://arxiv.org/abs/1807.07003}{{\ttfamily
  1807.07003}}].

\bibitem{Coriano:2018tgn}
C.~Corian\`o and M.~M. Maglio, \emph{{Conformal Ward Identities and the
  Coupling of QED and QCD to Gravity}},
  \href{https://doi.org/10.1051/epjconf/201819200047}{\emph{EPJ Web Conf.}
  {\bfseries 192} (2018) 00047}
  [\href{https://arxiv.org/abs/1809.05940}{{\ttfamily 1809.05940}}].

\bibitem{Albayrak:2018tam}
S.~Albayrak and S.~Kharel, \emph{{Towards the higher point holographic momentum
  space amplitudes}},
  \href{https://doi.org/10.1007/JHEP02(2019)040}{\emph{JHEP} {\bfseries 02}
  (2019) 040} [\href{https://arxiv.org/abs/1810.12459}{{\ttfamily
  1810.12459}}].

\bibitem{Farrow:2018yni}
J.~A. Farrow, A.~E. Lipstein and P.~McFadden, \emph{{Double copy structure of
  CFT correlators}}, \href{https://doi.org/10.1007/JHEP02(2019)130}{\emph{JHEP}
  {\bfseries 02} (2019) 130}
  [\href{https://arxiv.org/abs/1812.11129}{{\ttfamily 1812.11129}}].

\bibitem{Isono:2018rrb}
H.~Isono, T.~Noumi and G.~Shiu, \emph{{Momentum space approach to crossing
  symmetric CFT correlators}},
  \href{https://doi.org/10.1007/JHEP07(2018)136}{\emph{JHEP} {\bfseries 07}
  (2018) 136} [\href{https://arxiv.org/abs/1805.11107}{{\ttfamily
  1805.11107}}].

\bibitem{Isono:2019wex}
H.~Isono, T.~Noumi and G.~Shiu, \emph{{Momentum space approach to crossing
  symmetric CFT correlators. Part II. General spacetime dimension}},
  \href{https://doi.org/10.1007/JHEP10(2019)183}{\emph{JHEP} {\bfseries 10}
  (2019) 183} [\href{https://arxiv.org/abs/1908.04572}{{\ttfamily
  1908.04572}}].

\bibitem{Isono:2019ihz}
H.~Isono, T.~Noumi and T.~Takeuchi, \emph{{Momentum space conformal three-point
  functions of conserved currents and a general spinning operator}},
  \href{https://doi.org/10.1007/JHEP05(2019)057}{\emph{JHEP} {\bfseries 05}
  (2019) 057} [\href{https://arxiv.org/abs/1903.01110}{{\ttfamily
  1903.01110}}].

\bibitem{Maglio:2019grh}
C.~Corian\`o and M.~M. Maglio, \emph{{On Some Hypergeometric Solutions of the
  Conformal Ward Identities of Scalar 4-point Functions in Momentum Space}},
  \href{https://doi.org/10.1007/JHEP09(2019)107}{\emph{JHEP} {\bfseries 09}
  (2019) 107} [\href{https://arxiv.org/abs/1903.05047}{{\ttfamily
  1903.05047}}].

\bibitem{Gillioz:2019lgs}
M.~Gillioz, \emph{{Conformal 3-point functions and the Lorentzian OPE in
  momentum space}},
  \href{https://doi.org/10.1007/s00220-020-03836-8}{\emph{Commun. Math. Phys.}
  {\bfseries 379} (2020) 227}
  [\href{https://arxiv.org/abs/1909.00878}{{\ttfamily 1909.00878}}].

\bibitem{Bzowski:2019kwd}
A.~Bzowski, P.~McFadden and K.~Skenderis, \emph{{Conformal $n$-point functions
  in momentum space}},
  \href{https://doi.org/10.1103/PhysRevLett.124.131602}{\emph{Phys. Rev. Lett.}
  {\bfseries 124} (2020) 131602}
  [\href{https://arxiv.org/abs/1910.10162}{{\ttfamily 1910.10162}}].

\bibitem{Gillioz:2019iye}
M.~Gillioz, X.~Lu, M.~A. Luty and G.~Mikaberidze, \emph{{Convergent
  Momentum-Space OPE and Bootstrap Equations in Conformal Field Theory}},
  \href{https://doi.org/10.1007/JHEP03(2020)102}{\emph{JHEP} {\bfseries 03}
  (2020) 102} [\href{https://arxiv.org/abs/1912.05550}{{\ttfamily
  1912.05550}}].

\bibitem{Bautista:2019qxj}
T.~Bautista and H.~Godazgar, \emph{{Lorentzian CFT 3-point functions in
  momentum space}}, \href{https://doi.org/10.1007/JHEP01(2020)142}{\emph{JHEP}
  {\bfseries 01} (2020) 142}
  [\href{https://arxiv.org/abs/1908.04733}{{\ttfamily 1908.04733}}].

\bibitem{Coriano:2019nkw}
C.~Corian\`o, M.~M. Maglio and D.~Theofilopoulos, \emph{{Four-Point Functions
  in Momentum Space: Conformal Ward Identities in the Scalar/Tensor case}},
  \href{https://doi.org/10.1140/epjc/s10052-020-8089-1}{\emph{Eur. Phys. J. C}
  {\bfseries 80} (2020) 540}
  [\href{https://arxiv.org/abs/1912.01907}{{\ttfamily 1912.01907}}].

\bibitem{Lipstein:2019mpu}
A.~E. Lipstein and P.~McFadden, \emph{{Double copy structure and the flat space
  limit of conformal correlators in even dimensions}},
  \href{https://doi.org/10.1103/PhysRevD.101.125006}{\emph{Phys. Rev. D}
  {\bfseries 101} (2020) 125006}
  [\href{https://arxiv.org/abs/1912.10046}{{\ttfamily 1912.10046}}].

\bibitem{Bzowski:2020kfw}
A.~Bzowski, P.~McFadden and K.~Skenderis, \emph{{Conformal correlators as
  simplex integrals in momentum space}},
  \href{https://doi.org/10.1007/JHEP01(2021)192}{\emph{JHEP} {\bfseries 01}
  (2021) 192} [\href{https://arxiv.org/abs/2008.07543}{{\ttfamily
  2008.07543}}].

\bibitem{Jain:2020rmw}
S.~Jain, R.~R. John and V.~Malvimat, \emph{{Momentum space spinning correlators
  and higher spin equations in three dimensions}},
  \href{https://doi.org/10.1007/JHEP11(2020)049}{\emph{JHEP} {\bfseries 11}
  (2020) 049} [\href{https://arxiv.org/abs/2005.07212}{{\ttfamily
  2005.07212}}].

\bibitem{Jain:2020puw}
S.~Jain, R.~R. John and V.~Malvimat, \emph{{Constraining momentum space
  correlators using slightly broken higher spin symmetry}},
  \href{https://arxiv.org/abs/2008.08610}{{\ttfamily 2008.08610}}.

\bibitem{Coriano:2020ccb}
C.~Corian\`o and M.~M. Maglio, \emph{{The Generalized Hypergeometric Structure
  of the Ward Identities of CFT\textquoteright{}s in Momentum Space in $d >
  2$}}, \href{https://doi.org/10.3390/axioms9020054}{\emph{Axioms} {\bfseries
  9} (2020) 54} [\href{https://arxiv.org/abs/2001.09622}{{\ttfamily
  2001.09622}}].

\bibitem{Albayrak:2020fyp}
S.~Albayrak, S.~Kharel and D.~Meltzer, \emph{{On duality of color and
  kinematics in (A)dS momentum space}},
  \href{https://doi.org/10.1007/JHEP03(2021)249}{\emph{JHEP} {\bfseries 03}
  (2021) 249} [\href{https://arxiv.org/abs/2012.10460}{{\ttfamily
  2012.10460}}].

\bibitem{Armstrong:2020woi}
C.~Armstrong, A.~E. Lipstein and J.~Mei, \emph{{Color/kinematics duality in
  AdS$_{4}$}}, \href{https://doi.org/10.1007/JHEP02(2021)194}{\emph{JHEP}
  {\bfseries 02} (2021) 194}
  [\href{https://arxiv.org/abs/2012.02059}{{\ttfamily 2012.02059}}].

\bibitem{Serino:2020pyu}
M.~Serino, \emph{{The four-point correlation function of the energy-momentum
  tensor in the free conformal field theory of a scalar field}},
  \href{https://doi.org/10.1140/epjc/s10052-020-8208-z}{\emph{Eur. Phys. J. C}
  {\bfseries 80} (2020) 686}
  [\href{https://arxiv.org/abs/2004.08668}{{\ttfamily 2004.08668}}].

\bibitem{Coriano:2020ees}
C.~Corian\`o and M.~M. Maglio, \emph{{Conformal Field Theory in Momentum Space
  and Anomaly Actions in Gravity: The Analysis of 3- and 4-Point Functions}},
  \href{https://arxiv.org/abs/2005.06873}{{\ttfamily 2005.06873}}.

\bibitem{Mata:2012bx}
I.~Mata, S.~Raju and S.~Trivedi, \emph{{CMB from CFT}},
  \href{https://doi.org/10.1007/JHEP07(2013)015}{\emph{JHEP} {\bfseries 07}
  (2013) 015} [\href{https://arxiv.org/abs/1211.5482}{{\ttfamily 1211.5482}}].

\bibitem{Ghosh:2014kba}
A.~Ghosh, N.~Kundu, S.~Raju and S.~P. Trivedi, \emph{{Conformal Invariance and
  the Four Point Scalar Correlator in Slow-Roll Inflation}},
  \href{https://doi.org/10.1007/JHEP07(2014)011}{\emph{JHEP} {\bfseries 07}
  (2014) 011} [\href{https://arxiv.org/abs/1401.1426}{{\ttfamily 1401.1426}}].

\bibitem{Kundu:2014gxa}
N.~Kundu, A.~Shukla and S.~P. Trivedi, \emph{{Constraints from Conformal
  Symmetry on the Three Point Scalar Correlator in Inflation}},
  \href{https://doi.org/10.1007/JHEP04(2015)061}{\emph{JHEP} {\bfseries 04}
  (2015) 061} [\href{https://arxiv.org/abs/1410.2606}{{\ttfamily 1410.2606}}].

\bibitem{Arkani-Hamed:2015bza}
N.~Arkani-Hamed and J.~Maldacena, \emph{{Cosmological Collider Physics}},
  \href{https://arxiv.org/abs/1503.08043}{{\ttfamily 1503.08043}}.

\bibitem{Maldacena:2011nz}
J.~M. Maldacena and G.~L. Pimentel, \emph{{On graviton non-Gaussianities during
  inflation}}, \href{https://doi.org/10.1007/JHEP09(2011)045}{\emph{JHEP}
  {\bfseries 09} (2011) 045} [\href{https://arxiv.org/abs/1104.2846}{{\ttfamily
  1104.2846}}].

\bibitem{Arkani-Hamed:2018kmz}
N.~Arkani-Hamed, D.~Baumann, H.~Lee and G.~L. Pimentel, \emph{{The Cosmological
  Bootstrap: Inflationary Correlators from Symmetries and Singularities}},
  \href{https://doi.org/10.1007/JHEP04(2020)105}{\emph{JHEP} {\bfseries 04}
  (2020) 105} [\href{https://arxiv.org/abs/1811.00024}{{\ttfamily
  1811.00024}}].

\bibitem{Baumann:2019oyu}
D.~Baumann, C.~Duaso~Pueyo, A.~Joyce, H.~Lee and G.~L. Pimentel, \emph{{The
  cosmological bootstrap: weight-shifting operators and scalar seeds}},
  \href{https://doi.org/10.1007/JHEP12(2020)204}{\emph{JHEP} {\bfseries 12}
  (2020) 204} [\href{https://arxiv.org/abs/1910.14051}{{\ttfamily
  1910.14051}}].

\bibitem{Baumann:2020dch}
D.~Baumann, C.~Duaso~Pueyo, A.~Joyce, H.~Lee and G.~L. Pimentel, \emph{{The
  Cosmological Bootstrap: Spinning Correlators from Symmetries and
  Factorization}},  \href{https://arxiv.org/abs/2005.04234}{{\ttfamily
  2005.04234}}.

\bibitem{Skvortsov:2018uru}
E.~Skvortsov, \emph{{Light-Front Bootstrap for Chern-Simons Matter Theories}},
  \href{https://doi.org/10.1007/JHEP06(2019)058}{\emph{JHEP} {\bfseries 06}
  (2019) 058} [\href{https://arxiv.org/abs/1811.12333}{{\ttfamily
  1811.12333}}].

\bibitem{Jain:2021wyn}
S.~Jain, R.~R. John, A.~Mehta, A.~A. Nizami and A.~Suresh, \emph{{Momentum
  space parity-odd CFT 3-point functions}},
  \href{https://arxiv.org/abs/2101.11635}{{\ttfamily 2101.11635}}.

\bibitem{Maldacena:2011jn}
J.~Maldacena and A.~Zhiboedov, \emph{{Constraining Conformal Field Theories
  with A Higher Spin Symmetry}},
  \href{https://doi.org/10.1088/1751-8113/46/21/214011}{\emph{J. Phys. A}
  {\bfseries 46} (2013) 214011}
  [\href{https://arxiv.org/abs/1112.1016}{{\ttfamily 1112.1016}}].

\bibitem{Maldacena:2012sf}
J.~Maldacena and A.~Zhiboedov, \emph{{Constraining conformal field theories
  with a slightly broken higher spin symmetry}},
  \href{https://doi.org/10.1088/0264-9381/30/10/104003}{\emph{Class. Quant.
  Grav.} {\bfseries 30} (2013) 104003}
  [\href{https://arxiv.org/abs/1204.3882}{{\ttfamily 1204.3882}}].

\bibitem{Giombi:2011rz}
S.~Giombi, S.~Prakash and X.~Yin, \emph{{A Note on CFT Correlators in Three
  Dimensions}}, \href{https://doi.org/10.1007/JHEP07(2013)105}{\emph{JHEP}
  {\bfseries 07} (2013) 105} [\href{https://arxiv.org/abs/1104.4317}{{\ttfamily
  1104.4317}}].

\bibitem{Aharony:2011jz}
O.~Aharony, G.~Gur-Ari and R.~Yacoby, \emph{{d=3 Bosonic Vector Models Coupled
  to Chern-Simons Gauge Theories}},
  \href{https://doi.org/10.1007/JHEP03(2012)037}{\emph{JHEP} {\bfseries 03}
  (2012) 037} [\href{https://arxiv.org/abs/1110.4382}{{\ttfamily 1110.4382}}].

\bibitem{Giombi:2011kc}
S.~Giombi, S.~Minwalla, S.~Prakash, S.~P. Trivedi, S.~R. Wadia and X.~Yin,
  \emph{{Chern-Simons Theory with Vector Fermion Matter}},
  \href{https://doi.org/10.1140/epjc/s10052-012-2112-0}{\emph{Eur. Phys. J. C}
  {\bfseries 72} (2012) 2112}
  [\href{https://arxiv.org/abs/1110.4386}{{\ttfamily 1110.4386}}].

\bibitem{Aharony:2012nh}
O.~Aharony, G.~Gur-Ari and R.~Yacoby, \emph{{Correlation Functions of Large N
  Chern-Simons-Matter Theories and Bosonization in Three Dimensions}},
  \href{https://doi.org/10.1007/JHEP12(2012)028}{\emph{JHEP} {\bfseries 12}
  (2012) 028} [\href{https://arxiv.org/abs/1207.4593}{{\ttfamily 1207.4593}}].

\bibitem{Giombi_2017}
S.~Giombi, V.~Gurucharan, V.~Kirilin, S.~Prakash and E.~Skvortsov, \emph{On the
  higher-spin spectrum in large n chern-simons vector models},
  \href{https://doi.org/10.1007/jhep01(2017)058}{\emph{Journal of High Energy
  Physics} {\bfseries 2017} (2017) }.

\bibitem{Chowdhury:2017vel}
S.~D. Chowdhury, J.~R. David and S.~Prakash, \emph{{Constraints on parity
  violating conformal field theories in $d=3$}},
  \href{https://doi.org/10.1007/JHEP11(2017)171}{\emph{JHEP} {\bfseries 11}
  (2017) 171} [\href{https://arxiv.org/abs/1707.03007}{{\ttfamily
  1707.03007}}].

\bibitem{Skvortsov_2019}
E.~Skvortsov, \emph{Light-front bootstrap for chern-simons matter theories},
  \href{https://doi.org/10.1007/jhep06(2019)058}{\emph{Journal of High Energy
  Physics} {\bfseries 2019} (2019) }.

\bibitem{wip}
\emph{{To appear soon}}, .

\bibitem{Karateev:2017jgd}
D.~Karateev, P.~Kravchuk and D.~Simmons-Duffin, \emph{{Weight Shifting
  Operators and Conformal Blocks}},
  \href{https://doi.org/10.1007/JHEP02(2018)081}{\emph{JHEP} {\bfseries 02}
  (2018) 081} [\href{https://arxiv.org/abs/1706.07813}{{\ttfamily
  1706.07813}}].

\bibitem{Dobrev:1975ru}
V.~K. Dobrev, V.~B. Petkova, S.~G. Petrova and I.~T. Todorov, \emph{{Dynamical
  Derivation of Vacuum Operator Product Expansion in Euclidean Conformal
  Quantum Field Theory}},
  \href{https://doi.org/10.1103/PhysRevD.13.887}{\emph{Phys. Rev. D} {\bfseries
  13} (1976) 887}.

\end{thebibliography}\endgroup
\bibliographystyle{JHEP}
\end{document}